\documentclass[superscriptaddress,prc,twocolumn,showpacs,preprintnumbers,altaffilletter,aps,10pt,nofootinbib]{revtex4-1}
\usepackage{amssymb}
\usepackage{graphicx}
\usepackage{setspace}
\usepackage[caption=false]{subfig}
\usepackage{lipsum}
\usepackage{multirow}
\usepackage{siunitx}
\usepackage{lineno}
\usepackage{verbatim}
\usepackage{datetime} %MF. To include the current time in the date stamp.
\usepackage{xspace}
\usepackage{hyphenat}
\usepackage{bbold}
\usepackage{mathtools} %loads amsmath as well
\usepackage{ulem}
\usepackage{enumitem}

\newcommand{\mgcom}[1]{\textcolor{black}{#1}}

\usepackage{hyperref}
\usepackage{color}
\definecolor{linkcol}{rgb}{0,0,0.4} 
\definecolor{citecol}{rgb}{0.5,0,0} 
\hypersetup
{
pdftitle="Bayesian Analysis of a Future Beta Decay Experiment's Sensitivity to Neutrino Mass Scale and Ordering",
bookmarks=true,         % show bookmarks bar?
unicode=false,          % non-Latin characters in Acrobats bookmarks
pdftoolbar=true,        % show Acrobat toolbar?
pdfmenubar=true,        % show Acrobat menu?
pdffitwindow=true,     % window fit to page when opened
pdfstartview={FitH},    % fits the width of the page to the window
colorlinks=true,       % false: boxed links; true: colored links
linkcolor=blue,          % color of internal links
citecolor=blue,        % color of links to bibliography
}

\begin{document}
\title{Bayesian Analysis of a Future Beta Decay Experiment's Sensitivity to Neutrino Mass Scale and Ordering} 
\author{A.~Ashtari Esfahani} 
\affiliation{Center for Experimental Nuclear Physics and Astrophysics and Department of Physics, University of Washington, Seattle, WA 98195, USA}
\author{M.~Betancourt}
\affiliation{Symplectomorphic, LLC, New York, NY 10026, USA}
\author{Z.~Bogorad}\affiliation{Laboratory for Nuclear Science, Massachusetts Institute of Technology, Cambridge, MA 02139, USA}
\author{S.~B\"oser} \affiliation{Institut f\"ur Physik, Johannes Gutenberg-Universit\"at Mainz, 55128 Mainz, Germany}
\author{N.~Buzinsky} \affiliation{Laboratory for Nuclear Science, Massachusetts Institute of Technology, Cambridge, MA 02139, USA}
\author{R.~Cervantes} \affiliation{Center for Experimental Nuclear Physics and Astrophysics and Department of Physics, University of Washington, Seattle, WA 98195, USA}
\author{C.~Claessens} \affiliation{Institut f\"ur Physik, Johannes Gutenberg-Universit\"at Mainz, 55128 Mainz, Germany}
\author{L.~de~Viveiros} \affiliation{Department of Physics, Pennsylvania State University, University Park, PA 16802, USA}
%\author{S.~Doeleman} \affiliation{Harvard-Smithsonian Center for Astrophysics, Cambridge, MA 02138, USA}
\author{M.~Fertl} \affiliation{Institut f\"ur Physik, Johannes Gutenberg-Universit\"at Mainz, 55128 Mainz, Germany}
\author{J.~A.~Formaggio} \affiliation{Laboratory for Nuclear Science, Massachusetts Institute of Technology, Cambridge, MA 02139, USA}
\author{L.~Gladstone} \affiliation{Department of Physics, Case Western Reserve University, Cleveland, OH 44106, USA}
%\author{M.~G\"odel}\affiliation{Institut f\"ur Physik, Johannes Gutenberg-Universit\"at Mainz, 55128 Mainz, Germany}
\author{M.~Grando} \affiliation{Pacific Northwest National Laboratory, Richland, WA 99354, USA} 
\author{M.~Guigue} \email{mguigue@lpnhe.in2p3.fr} \affiliation{Laboratoire de Physique Nucl\'eaire et de Hautes \'Energies, Sorbonne Universit\'e, Universit\'e de Paris, CNRS/IN2P3, Paris, France} % \affiliation{Pacific Northwest National Laboratory, Richland, WA 99354, USA} 
\author{J.~Hartse}\affiliation{Center for Experimental Nuclear Physics and Astrophysics and Department of Physics, University of Washington, Seattle, WA 98195, USA}
\author{K.~M.~Heeger} \affiliation{Wright Laboratory, Department of Physics, Yale University, New Haven, CT 06520, USA}
\author{X.~Huyan} \affiliation{Pacific Northwest National Laboratory, Richland, WA 99354, USA} 
\author{J.~Johnston} \affiliation{Laboratory for Nuclear Science, Massachusetts Institute of Technology, Cambridge, MA 02139, USA}
\author{A.~M.~Jones} \affiliation{Pacific Northwest National Laboratory, Richland, WA 99354, USA}
\author{K.~Kazkaz} \affiliation{Lawrence Livermore National Laboratory, Livermore, CA 94550, USA}
\author{B.~H.~LaRoque} \affiliation{Pacific Northwest National Laboratory, Richland, WA 99354, USA}
\author{A.~Lindman} \affiliation{Institut f\"ur Physik, Johannes Gutenberg-Universit\"at Mainz, 55128 Mainz, Germany}
\author{R.~Mohiuddin} \affiliation{Department of Physics, Case Western Reserve University, Cleveland, OH 44106, USA}
\author{B.~Monreal} \affiliation{Department of Physics, Case Western Reserve University, Cleveland, OH 44106, USA}
\author{J.~A.~Nikkel} \affiliation{Wright Laboratory, Department of Physics, Yale University, New Haven, CT 06520, USA}
\author{E.~Novitski} \affiliation{Center for Experimental Nuclear Physics and Astrophysics and Department of Physics, University of Washington, Seattle, WA 98195, USA}
\author{N.~S.~Oblath} \affiliation{Pacific Northwest National Laboratory, Richland, WA 99354, USA}
\author{M.~Ottiger} \affiliation{Center for Experimental Nuclear Physics and Astrophysics and Department of Physics, University of Washington, Seattle, WA 98195, USA}
\author{W.~Pettus} \affiliation{Center for Experimental Nuclear Physics and Astrophysics and Department of Physics, University of Washington, Seattle, WA 98195, USA}
\affiliation{Department of Physics, Indiana University, Bloomington, IN 47405, USA}
\author{R.~G.~H.~Robertson} \affiliation{Center for Experimental Nuclear Physics and Astrophysics and Department of Physics, University of Washington, Seattle, WA 98195, USA}
\author{G.~ Rybka} \affiliation{Center for Experimental Nuclear Physics and Astrophysics and Department of Physics, University of Washington, Seattle, WA 98195, USA}
\author{L.~Salda\~na} \affiliation{Wright Laboratory, Department of Physics, Yale University, New Haven, CT 06520, USA}
\author{M.~Schram} \affiliation{Pacific Northwest National Laboratory, Richland, WA 99354, USA}
\author{V.~Sibille} \affiliation{Laboratory for Nuclear Science, Massachusetts Institute of Technology, Cambridge, MA 02139, USA}
\author{P.~L.~Slocum} \affiliation{Wright Laboratory, Department of Physics, Yale University, New Haven, CT 06520, USA}
\author{Y.-H.~Sun} \affiliation{Department of Physics, Case Western Reserve University, Cleveland, OH 44106, USA}
\author{P.~T.~Surukuchi}\affiliation{Wright Laboratory, Department of Physics, Yale University, New Haven, CT 06520, USA}
\author{J.~R.~Tedeschi} \affiliation{Pacific Northwest National Laboratory, Richland, WA 99354, USA}
\author{A.~B.~Telles}\affiliation{Wright Laboratory, Department of Physics, Yale University, New Haven, CT 06520, USA}
\author{M.~Thomas} \affiliation{Pacific Northwest National Laboratory, Richland, WA 99354, USA} 
\author{T.~Th\"ummler} \affiliation{Institute for Astroparticle Physics, Karlsruhe Institute of Technology, 76021 Karlsruhe, Germany}
\author{L.~Tvrznikova} \altaffiliation{Present address: Waymo LLC, Mountain View, CA 94043, USA} \affiliation{Lawrence Livermore National Laboratory, Livermore, CA 94550, USA}
\author{B.~A.~VanDevender} \affiliation{Center for Experimental Nuclear Physics and Astrophysics and Department of Physics, University of Washington, Seattle, WA 98195, USA} \affiliation{Pacific Northwest National Laboratory, Richland, WA 99354, USA}
%\author{J.~Weintroub} \affiliation{Harvard-Smithsonian Center for Astrophysics, Cambridge, MA 02138, USA}
\author{T.~E.~Weiss} \email{talia.weiss@yale.edu}
\affiliation{Laboratory for Nuclear Science, Massachusetts Institute of Technology, Cambridge, MA 02139, USA}
\affiliation{Wright Laboratory, Department of Physics, Yale University, New Haven, CT 06520, USA}
\author{T.~Wendler} \altaffiliation{Present address: Pacific Northwest National Laboratory, Richland, WA 99354, USA} \affiliation{Department of Physics, Pennsylvania State University, University Park, PA 16802, USA}
\author{E.~Zayas} \affiliation{Laboratory for Nuclear Science, Massachusetts Institute of Technology, Cambridge, MA 02139, USA}
\author{A.~Ziegler}\affiliation{Department of Physics, Pennsylvania State University, University Park, PA 16802, USA}

\begin{abstract}
Bayesian modeling techniques enable sensitivity analyses that incorporate detailed expectations regarding future experiments. A model-based approach also allows one to evaluate inferences and predicted outcomes, by calibrating (or measuring) the consequences incurred when certain results are reported. We present procedures for calibrating predictions of an experiment's sensitivity to both continuous and discrete parameters. Using these procedures and a new Bayesian model of the $\beta$-decay spectrum, we assess a high-precision $\beta$-decay experiment's sensitivity to the neutrino mass scale and ordering, for one assumed design scenario. We find that such an experiment could measure the electron-weighted neutrino mass within $\sim40\,$meV after 1 year (90$\%$ credibility). Neutrino masses $>500\,$meV could be measured within $\approx5\,$meV. Using only $\beta$-decay and external reactor neutrino data, we find that next-generation $\beta$-decay experiments could potentially constrain the mass ordering using a two-neutrino spectral model analysis. By calibrating mass ordering results, we identify reporting criteria that can be tuned to suppress false ordering claims. In some cases, a two-neutrino analysis can reveal that the mass ordering is inverted, an unobtainable result for the traditional one-neutrino analysis approach.
\end{abstract}
\maketitle

\section{Introduction}\label{I}
%\linenumbers
Model-based simulation is a standard tool for informing the design of physics experiments and predicting their outcomes~\cite{PDG2020}.  Such model-based approaches allow one to incorporate detailed expectations regarding future data by performing pseudo-experiments that reflect the span of possible experimental and physical parameter values. In Bayesian sensitivity studies, specifically, those parameter values are weighted by prior probabilities. By contrast, computing and reporting predicted outcomes for ``best guess" values ignores information by excluding regions of parameter space.

Moreover, inferential models lend themselves to procedures for investigating the consequences of assumptions made during analysis. Bayesian methods, in particular, illuminate the effects of assumptions underlying inference (i.e., extracting information from data) and decision making (i.e., claiming results based on inferences) by decoupling the two processes. Thus, when assessing an experiment's sensitivity, one can quantify, or {\it calibrate}, the expected success or accuracy of  procedures that one plans to use to both analyze data and report results in a certain format. It is also possible to perform conditional Bayesian calibration by fixing one or more parameters before simulating data~\cite{Betancourt2018, Little2006, Sellke2001, LacosteJulien2011}.
%, can enable rigorous studies of experimental sensitivity to quantities of interest.

\begin{table*}[t]
\begin{center}
\begin{tabular}{| l || l | l |}
\hline
\textbf{Term} & \textbf{Definition} & \textbf{Notes}
\tabularnewline
\hline
\hline
Credibility & Fraction of Bayesian posterior probability mass that falls  & Result of a {\it single} real or  \\
 &  within a reported interval  & simulated experiment \\
 \hline
 Coverage & Fraction of likely experiments for which the reported interval & Result of {\it multiple}  \\
 & contains the true parameter value, within model assumptions & simulated experiments \\
 \hline
 Confidence interval &  Interval constructed to have a coverage that equals or exceeds & Frequentist term; not used \\
 &  a chosen probability (or ``confidence level") & in this analysis \\
 \hline
 Sensitivity analysis & Study of how result precision \& accuracy change under reaso-  & Requires simulated \\
 & nable variation of all parameters, within model assumptions &  experiments (pseudo-data) \\
 \hline
  Sensitivity {\it(to very}  & Upper limit on a parameter, to some confidence level & Usage by the KATRIN \\
 {\it small parameter)} & & experiment~\cite{KATRIN2004} \\
 \hline
 Sensitivity {\it(to parameter}  & Width of a posterior interval with a chosen credibility & Definition in this paper \\
  {\it of any magnitude)} &  &  \\
\hline
\end{tabular}
\end{center}
\vspace{-0.5cm}
\caption{Definitions are consistent with Particle Data Group descriptions~\cite{PDG2020} with the exception of the two definitions of ``sensitivity," which capture a common but less standard usage. The last row describes how ``sensitivity" is used in this paper.}
\label{tab:StatsTerms}
\end{table*}

Here, we employ Bayesian modeling to perform a sensitivity study for a physics experiment.
%that is, to evaluate how precisely and accurately experimenters can expect to measure parameters of interest.
Among physicists, {\it sensitivity} typically denotes the level of precision with which experimenters can expect to resolve a parameter of interest, assuming a reasonably accurate measurement. (We adopt that usage here, though among statisticians, sensitivity can refer to how a decision making process' accuracy depends on model parameters~\cite{Betancourt2018, Little2006}.) For physics experiments, in particular, Bayesian sensitivity methods allow researchers to capitalize on their often extensive knowledge of experimental configurations, physical processes, and expected uncertainties to construct priors. More broadly, model-based analyses offer potential tools for physicists to collectively interpret results and judge whether discovery claims are warranted~\cite{Gustafson2009, Betancourt2018} (see Section~\ref{II}). These tools thus provide possible alternatives to a $5\sigma$ confidence requirement.

To assess sensitivity, we develop a model of an experiment's measurement process, then employ that model to repeatedly generate and analyze pseudo-data---where ``analyze" means ``infer posterior distributions." Parameters assumed for data generation are sampled from priors. Next, expectations and intervals are computed from the posteriors, yielding sensitivity results. Finally, we calculate how often these results reflect ``true" values underlying the generated data (a calibration). In doing so, we quantify the consequences of our modeling and reporting assumptions. For relevant statistical term definitions, see Table~\ref{tab:StatsTerms}.

The above procedure is applied here to assess sensitivity to the neutrino mass scale and ordering.
Neutrinos are produced in one of three flavor states, each of which interacts with electrons, muons or tau leptons. The discovery of neutrino oscillations demonstrated that each flavor state can be represented as a superposition of mass states with eigenvalues $m_1$, $m_2$ and $m_3$, at least two of which are nonzero~\cite{Abe2011, Aharmin2010, Eguchi2003}. While nuclear and particle physics experiments as well as cosmological models have placed upper bounds on the masses and measured the squared mass differences~\cite{PDG2020}, the absolute neutrino mass scale is unknown. In addition, two orderings of the mass spectrum are possible:  if $m_1 < m_2 < m_3$, the masses are said to obey a {\it normal ordering,} while if $m_3 < m_1 < m_2$, they follow an {\it inverted ordering}. Although recent data are beginning to shed light on the ordering question, it remains unanswered to date.
Sensitivity to the ordering in oscillation experiments is discussed in Qian {\it et al.}~\cite{Qian2012}.

A promising approach to resolving the mass scale involves analyzing the shape of the electron spectrum produced when nuclei $\beta$-decay. This ``direct mass measurement'' method is so named because it depends chiefly on decay kinematics imposed by energy conservation. Direct mass experiments probe the electron-weighted neutrino mass $m_\beta = \sqrt{ \sum_{i=1}^3 |U_{ei}|^2 m_i^2}$ (hereafter ``neutrino mass"), where \unexpanded{$U_{ei}$} are Pontecorvo-Maki-Nakagawa-Sakata (PMNS) matrix elements.\footnote{For either neutrino mass ordering, $m_\beta=m_1$ to 1\% accuracy for $m_1\gtrapprox0.05$\,eV. Hence, with knowledge of the ordering and splittings, an $m_\beta$ measurement determines all three masses.}
The size of $m_\beta$ corresponds to a shift in the electron's maximal kinetic energy and causes a distortion in the $\beta$ spectrum shape.
%as discussed in Section \ref{sec:one-two-model}.
A precise $m_\beta$ measurement would determine the mass scale, {and as a by-product, it could constrain the ordering at masses $\lessapprox$\,48\,meV---the 95\% lower limit on the inverted ordering mass~\cite{PDG2020}. % Zyla2020 
Furthermore, the $\beta$-decay shape depends distinctly on each $m_i$~\cite{Formaggio2014}. Thus, we propose that, if a $\beta$-decay experiment is sensitive to the fractional contributions of individual neutrino masses to the full spectral shape, then such information might enable a clearer mass ordering determination. By modeling the shape of a $\beta$ spectrum, one can thus assess a direct mass experiment's sensitivity to the ordering---accounting for both the magnitude of $m_\beta$ and finer spectral features (see Figure~\ref{fig:ordering-spectra}).

In this paper, we develop a $\beta$-decay spectral model suited to Bayesian inference. The model uses a two-neutrino approximation {(motivated by the fact that $\Delta m_{21}^2 \ll |\Delta m _{13}^2|$)} and formulates the mass ordering question in terms of a parameter $\eta$, the fractional contribution of the lighter mass to the spectrum. Constraints on $\eta$ are most directly accessible via reactor neutrino experiments. Thus, for a $\beta$-decay experiment to potentially resolve the mass ordering, the only external data needed for the analysis are  reactor data. Current as well as future direct mass experiments could employ this spectral model to examine their sensitivity to the neutrino mass scale and ordering. As a case study,
we use the model to assess sensitivity to these neutrino mass parameters for one possible design scenario of the Project 8 experiment, a high-precision $\beta$-decay experiment~\cite{Formaggio2009,Esfahani2017}.
%As a case study, we use the model to examine the potential sensitivity of a high-precision $\beta$-decay experiment---the Project 8 neutrino mass experiment---to those parameters~\cite{Formaggio2009,Esfahani2017}.

\begin{figure}[tb!]
\makebox[\linewidth][c]{\includegraphics[width=0.95\linewidth]{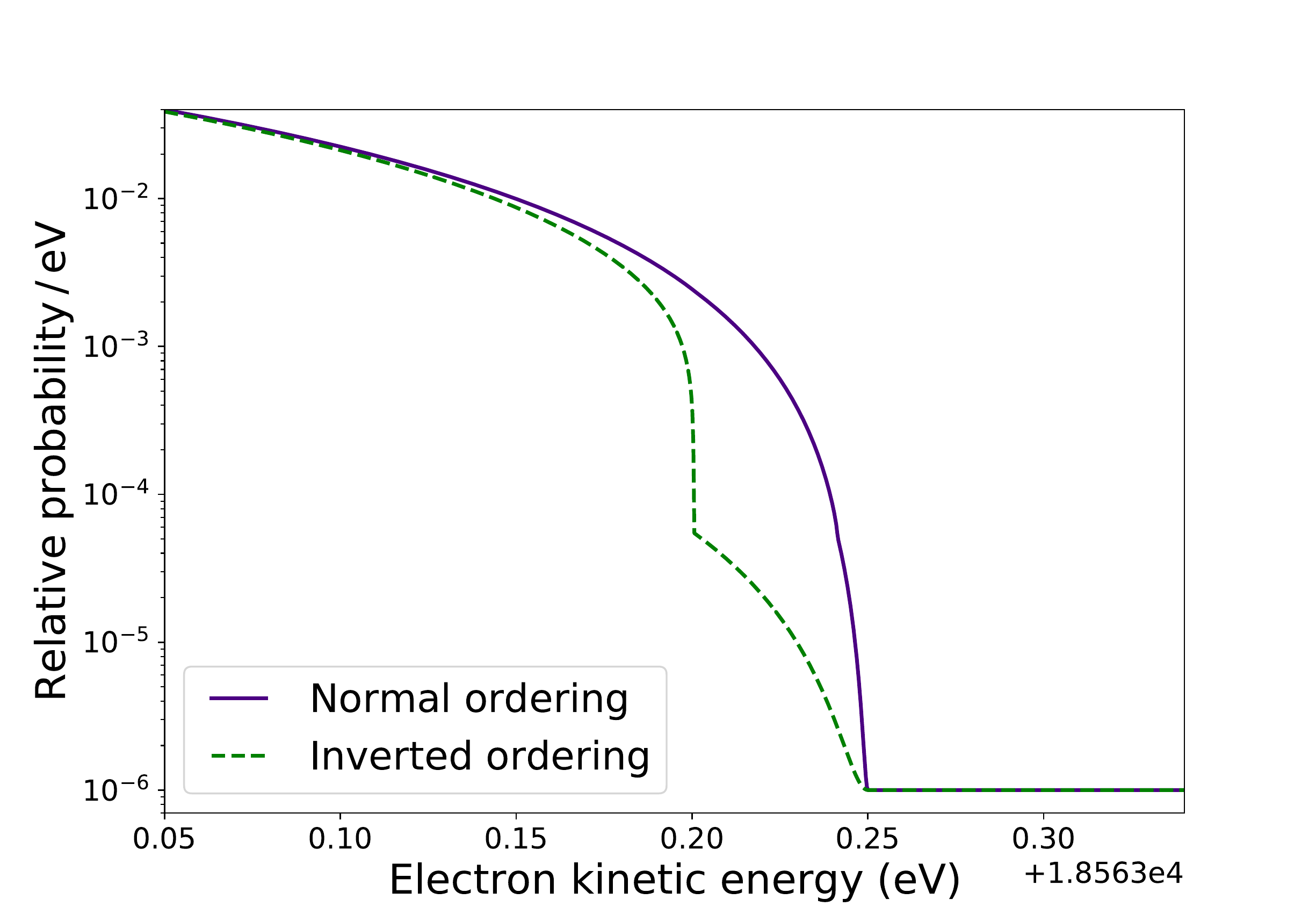}} %_ordering-mini-plot
\caption{A comparison of atomic tritium $\beta$-decay spectra for the two allowed neutrino mass orderings shows how a spectral shape analysis could be sensitive to the mass ordering. A background of $10^{-6}$/eV is assumed.}
\label{fig:ordering-spectra}
\end{figure}

\section{A Model-Based Approach to Calibrating Sensitivity Results}\label{II}
%Like a shortened version of chapter 2 of my thesis

Predictive analyses project whether, given some expected data, one will be able to report a particular result---for example, ``$m_\beta$ falls between 0.05 and 0.09 eV with 90\% credibility" or ``the
mass ordering is normal." In Bayesian analysis, the {\it decision} of whether to claim a particular result occurs after the process of {\it inference}. Bayesian inference produces posterior distributions $\pi(\theta | y)$ for parameters $\theta$ given data $y$. Such inference exploits Bayes' rule $\pi(\theta | y) \propto \pi(y | \theta) \cdot \pi(\theta),$ where $\pi(y | \theta)$ is the likelihood of $y$ given $\theta$, and $\pi(\theta)$ are prior probability distributions on $\theta$. Experimenters make claims about physics underlying their data by computing expectations (e.g. means and intervals) from posteriors.

In practice, there is no guarantee that the process by which one decides to claim a scientific result will perform well when faced
with real data. To provide some assurance of the decision making process' good performance, it is necessary to calibrate the process by evaluating it with respect to possible ``model configurations," i.e., combinations of true parameter values.

Decision-making procedures are, in this context, cost-benefit analyses. A calibration requires an inferential loss (or utility) function that expresses the relative loss $L$ incurred when an experimenter makes different reporting choices. In this work, $L$ lies between 0 and 1. A common choice for $L$ (used in Section~\ref{IIA}) is a function that equals 0 if a credible or confidence interval obtained by analyzing a pseudo-dataset contains the true parameter value, or 1 if it does not. The expected loss is then estimated by finding the average loss $\bar{L}$ for a group of pseudo-datasets. In the case just described, $\bar{L}$ would be the fraction of datasets for which the interval does {\it not} contain the true value. Given multiple reporting options, the experimenter should select the option with the smallest average loss over a group of pseudo-experiments~\cite{Gustafson2009, Betancourt2018}. (For example, this enables a decision of whether to report quantile or highest density intervals, as discussed further in Section~\ref{IIA}.)  

There is no one correct loss function for a given model, but the function should quantify the agreement or discrepancy between inputted and reported values. Given some loss function, model-based calibration then serves to compute how often one reports {\it accurate} results, across many pseudo-experiments with likely model configurations~\cite{Talts2018, Gustafson2009, Betancourt2018}.

%Decision-making procedures are, in this context, cost-benefit analyses. A calibration requires an inferential loss (or utility) function that expresses the relative loss $L$ incurred { when an experimenter makes} different reporting choices. Given multiple reporting options, that experimenter should select the option with the smallest  average loss over a group of pseudo-experiments There is no one correct loss function for a given model, but the function should quantify the agreement or discrepancy between inputted and reported values. Given some loss function, model-based calibration then involves computing how often one reports {\it accurate} results, across many pseudo-experiments with likely model configurations~\cite{ Talts2018, Gustafson2009, Betancourt2018}.

Frequentist calibration entails finding the {\it worst-case loss} over all model configurations. Such calibration requires tools like the Feldman-Cousins method, which addresses the fact that typical, Gaussian confidence intervals are inaccurate for bounded parameters, such as the positive neutrino mass~\cite{FeldmanCousins, Neyman1937}. This approach is too time-consuming to implement fully, as it requires that likelihood functions be computed and integrated for all reasonable parameter values (or a fine grid). While asymptotic approximations can make frequentist calibration computationally viable, they do not fully hold for the complex statistical models used in modern analyses~\cite{Berger2004, ModelingCaseStudy}.
%Such claims usually tend to be made depending on the outcome of the experiment: indeed, if the built confidence interval on a parameter of interest contains a natural boundary, e.g. 0, experimenters will be tempted to present their result as a limit of such parameter, leading to a flip-flop issue. One of the most commonly used and reliable method to build such confidence intervals is the Feldman-Cousins approach~\cite{FeldmanCousins}. This approach relies on the Neyman's construction of confidence intervals~\cite{Neyman1937} that resolves several issues including the ``flip-flopping" policy. However, this method can sometimes be time-consuming to implement, as it requires that one compute the probability of obtaining a given dataset as a function of parameters in the region of interest, before the result of the experiment can be compared to the built confidence belt.]

%By contrast, Bayesian calibration is computationally manageable, because it involves sampling model configurations from priors to examine the resulting distribution of losses.  Hence, in Bayesian analysis, it is not necessary to consider all possible truths---only those favored by priors
Bayesian calibration, on the other hand, does not require that one determine the {\it worst-case loss}; instead, it entails finding the {\it expected loss with respect to the prior distribution.} This is a probabilistic calculation that can be readily implemented with sampling methods. In a Bayesian analysis, it is not necessary to consider all possible truths---only enough to accurately estimate expected losses~\cite{ModelingCaseStudy}. Here, we lay out Bayesian calibration procedures for sensitivity to the electron-weighted neutrino mass and mass ordering.

\subsection{Calibrating Neutrino Mass Sensitivity Claims}\label{IIA}
The Bayesian result of a physics experiment will often be a posterior credible window---that is, the window within which some fraction of a parameter's posterior probability mass falls. This reporting scheme is sensible for continuous-domain parameters. If a posterior on $m_\beta$ is inferred from a $\beta$ spectrum, experimenters can present their result as a credible window of neutrino masses (in eV). We call the width of this window ``sensitivity to the neutrino mass." The reported mass window may consist of either an upper limit with a lower bound at zero, or a credible interval with upper and lower bounds. If posteriors are inferred for a large number of pseudo-data sets, one may predict an experiment's sensitivity by computing an expectation value (e.g. mean width or median width) from these credible windows. For a discussion of the frequentist and Bayesian perspectives underlying the use of confidence and credible intervals, respectively, see~\cite{Betancourt2018}.

In the continuous-domain case, the loss function provides a method for computing the proportion of likely data sets for which a posterior interval contains the true parameter value.
For an analysis of sensitivity to $m_\beta$, a calibration involves computing the fraction of pseudo-data sets $C \equiv 1-\bar{L}$ for which the credible window includes the true neutrino mass $\tilde{m}_\beta$, where $\bar{L}$ is the average loss for an ensemble of pseudo-experiments. ($\bar{L}$ serves to estimate the {\it expected loss} with respect to the prior distribution.) The fraction $C$ is known as the Bayesian ``model coverage," and it estimates the expected accuracy of a sensitivity prediction.\footnote{Note that credible intervals do not guarantee 
any frequentist coverage. Constructing confidence intervals and computing frequentist coverages would require analyzing an ensemble of pseudo-experiments for a multi-dimensional grid of input parameter configurations. This becomes impractical in many dimensions, where the number of configurations on any reasonably sized grid grows exponentially fast~\cite{Berger2004}.} We find $C$ by repeatedly generating and analyzing data given an appropriate distribution of inputted $\tilde{m}_\beta$ values~\cite{Betancourt2018}.

\vspace{7 pt}
The calibration procedure is as follows: 
\begin{enumerate}
\setlength\itemsep{-1 pt}
\item { Develop a {\it generation} model of the data, and if necessary, a second {\it analysis} model. The latter may be approximate but is believed to adequately describe the data. Both models depend on a set of parameters $\theta$ (which includes $m_\beta$).}
\item Select ``true" values $\tilde{\theta}$ by sampling from priors $\pi(\theta)$, which incorporate as much external knowledge as is reasonable.
\item Generate spectral data $\tilde{y}$ using { the generation model}, with $\tilde{\theta}$ as inputs.
\item Use the { analysis model} from \#1 to infer a posterior $\pi(m_\beta | \tilde{y})$.
\item Determine the posterior values $\vartheta$ that contain some fraction (credibility) $\alpha$ of the posterior probability mass on $m_\beta$. For a credible {\it interval}, calculate the loss function
\begin{equation}
L_{m_\beta} \equiv \begin{cases} 0, & \ \ \tilde{m}_\beta \in \Big[\vartheta_{(1-\alpha)/2}, \vartheta_{(1+\alpha)/2}\Big] \\
1, & \ \ \text{Otherwise},
\end{cases}
\label{eq:LossMbeta}
\end{equation}
\noindent where upper and lower posterior bounds $\vartheta_{(1\pm\alpha)/2}$ are computed so that
\begin{equation}
%\vartheta_\zeta = \text{the value of}\ \vartheta\ \text{for which}
\int_0^{\vartheta_{(1\pm\alpha)/2}} d m_\beta \, \pi(m_\beta | \tilde{y}) = \frac{1\pm\alpha}{2}.
\label{eq:CIdefinition}
\end{equation}
\noindent That is, a fraction $(1\pm\alpha)/2$ of the posterior probability mass on $m_\beta$ lies below the mass value $\vartheta_{(1\pm\alpha)/2}$. For a {\it limit}, the credible window is $[0, \vartheta_\alpha]$. 
\item Repeat steps 2--5 {$N_{\text{trial}}$ times. Each repetition constitutes a ``pseudo-experiment."} %$\sqrt{\frac{C(1-C)}{N_\text{exper}}}$.} 
\item Compute $C$ by subtracting the mean over resulting $L_{m_\beta}$ values from 1. Potentially, adjust $\alpha$ to obtain a satisfying coverage---that is, to achieve an acceptable number of true and false positive results.
\end{enumerate}
%$C =$ (number of times $\tilde{m}_\beta$ falls within the window)/(total number of pseudo-experiments).
\noindent $C$ may not equal $\alpha$ for all $\alpha$; the relationship between these two values depends on the model and priors. The uncertainty on $C$ is $\sqrt{C\cdot(1-C)/N_\text{trial}}$, assuming the number of true positive results is binomially distributed.

A calibrated sensitivity result then consists of a projected (e.g., mean or median) credible window and its coverage. {\it It is necessary to sample all input values from priors before generation} (step 2). This creates an ensemble of many realistic data sets, where the probabilities of possible model configurations are weighted appropriately. If a model-based sensitivity analysis uses fixed generation inputs (or a grid of inputs, unweighted by prior probabilities), it risks biasing results and under- or over-estimating coverages. {\it It is also crucial to generate pseudo-data that is as realistic as possible}, so that the coverage will reflect the potential consequences of all known assumptions made when devising the analysis model or choosing how to report results~\cite{Betancourt2018}.

Note that expected fluctuations in the data itself (i.e., statistical uncertainties) are incorporated into priors used for both data generation and analysis---steps 3 and 4. By contrast, uncertainties representing a lack of clarity in one's knowledge of fixed parameters (i.e., systematic uncertainties) are incorporated into pre-generation sampling and analysis priors---steps 2 and 4. 

In the above procedure, a choice of credibility $\alpha$ does not uniquely define a credible window, because one must also select the window’s central value. A straightforward choice of window is the quantile interval, which contains an equal amount of probability mass above and below the posterior median (as in Eqs.~\ref{eq:LossMbeta} and~\ref{eq:CIdefinition}). {For asymmetric posteriors, however, highest density intervals (HDIs) may be preferable. An HDI is computed by finding all credible intervals for a given $\alpha$ and selecting the narrowest interval. For a continuous posterior, this is equivalent to lowering a horizontal line over the posterior until the outermost intersection points between the line and curve contain a fraction $\alpha$ of posterior probability mass~\cite{Hyndman1996}. For a particular ensemble of posteriors, assuming both of these interval types are qualitatively sensible, one can decide which to adopt by computing and comparing coverages for each.}
%For the neutrino mass case, in Section~\ref{4a}, we employ the type of credible interval that yields the higher coverage. %calculate coverages using both approach types and report the one with higher $C$.}
%for which an equal fraction of the contained posterior mass falls on either side of the posterior mean.

{When measuring a continuous parameter like $m_\beta$, physicists are often concerned not only with  precision, but also with ``discovery potential": the probability that the parameter is nonzero. While neutrinos have been found to be massive \mgcom{through oscillation experiments}, a beta-decay result distinguishing $m_\beta$ from zero with high confidence or credibility would provide strong verification of physicists' interpretation of \mgcom{these} oscillation data~\cite{KATRIN2004}. Here, we claim a continuous parameter is nonzero if its highest density credible interval does not intersect with zero. (In practice, the $m_\beta$ prior affects the outcome; see Section~\ref{nonzero-mass}.)
% In practice, that procedure requires an adjustment due to the posteriors' discrete nature and the $m_\beta$ prior's effect;
%oscillation results demonstrate that

To verify that a scheme for assessing discovery potential is sound, a second calibration is required. This involves inputting a ``true" mass value of zero for an ensemble of pseudo-experiments, then constructing HDIs with some credibility. Next, one confirms that the interval credibility approximately equals the fraction (coverage) of experiments for which the interval contains zero.}

\subsection{Calibrating Mass Ordering Sensitivity Claims}\label{IIB}
It is similarly possible to calibrate the process of claiming that the neutrino masses obey one ordering. This process is an example of result reporting for a discrete-domain parameter. In that case, we follow the above procedure through step 4, replacing $m_\beta$ with a parameter that encodes mass ordering information. For our $\beta$ spectral model, that parameter is $\eta$, the lighter mass' contribution to the spectrum. For normal and inverted orderings, respectively, $\eta$ tends toward precisely known values $\eta_{N}$ and $\eta_{I}$  (see Section III). We claim a hypothetical ordering result when the posterior $\pi(\eta | \tilde{y})$ clusters near the predicted value for one ordering.

Specifically, as a suggested decision making scheme, we report a normal (inverted) ordering result when a posterior interval on $\eta$ with credibility $\kappa$ contains $\eta_N$ ($\eta_I$) but not $\eta_I$ ($\eta_N$). For a credible interval $T$ on $\eta$, the associated loss functions for each ordering are
\begin{equation}
\begin{split}
L_{N} &\equiv \begin{cases} 0, & \ \ (\eta_N \in T) \text{\ and\ } (\eta_I \notin T) \\
1, & \ \ \text{Otherwise}
\end{cases}\\
L_{I} &\equiv \begin{cases} 0, & \ \ (\eta_I \in T) \text{\ and\ } (\eta_N \notin T)  \\
1, & \ \ \text{Otherwise} \end{cases} \\
T & = \Big[\phi_{(1-\kappa)/2}, \phi_{(1+\kappa)/2}\Big],
\end{split}
\label{eq:Loss-ordering}
\end{equation}
\noindent where posterior bounds on $\eta$ are computed so that
\begin{equation}
\int_0^{\phi_{(1\pm\kappa)/2}} d\eta \, \pi(\eta | \tilde{y}) = \frac{1\pm\kappa}{2} .
\nonumber
\end{equation}

\noindent {These bounds may be selected using either a quantile or a highest density approach, depending on which yields higher coverage.} If $L_N=L_I=0$ or 1, neither ordering is strongly favored and nothing can be claimed.

%These true and false claim rates are affected by the credibility $\kappa$.
For each ``true" mass ordering, given a series of pseudo-experiments, we then compute the rates at which we report correct and incorrect mass ordering results (see Section~\ref{sec:ordering-results}). %(see Table~\ref{tab:mHsensitivity}).
These true and false claim rates enable experimenters to select a credibility $\kappa$---i.e., to decide how stringent to make their reporting criterion. As in the continuous parameter case, this calibration of sensitivity to the mass ordering should be performed for a large number of model configurations sampled from priors. {A similar calibration procedure would apply to accelerator, atmospheric and reactor experiments seeking to resolve the ordering~\cite{Qian2015}}, given an $\eta$-like parameter expressing mass ordering information.

We implement the above two procedures using the Stan software platform for Bayesian inference, which estimates posteriors by exploring a probability density parameter space using Markov Chain Monte Carlo methods (specifically, Hamiltonian Monte Carlo~\cite{Neal2011, Betancourt2015}). Stan is a valuable predictive analysis tool because it deals well with high dimensional problems and allows users to focus on modeling systems instead of developing computational architecture~\cite{Carpenter2017,Stanual2020}. Along with Stan, we employ morpho, a python-based tool we developed to organize information {input to and output from} Stan. Morpho facilitates a Stan workflow involving convergence checks and { analysis of posteriors}, and it is designed to suit general Stan users~\cite{morpho}.

\section{Model Formalism for a Beta Decay Experiment}
%(Based on Chapter 3 of and Eq.~4.1-4.2 in my thesis.) \\

The differential spectrum predicted for beta decay has a well understood analytic distribution, especially for superallowed transitions. The rate at which electrons are ejected as a function of their total energies is described by the equation
\begin{equation}
\begin{aligned}
\frac{dN}{dE_e} = &  \, \Bigg[\frac{G_F^2 |V_{ud}|^2}{2 \pi^3} |M_{\rm nuc}|^2 F(Z,p_e) p_e E_e \Bigg] \times \\
& \Bigg[ \sum_{i=1}^3 |U_{\rm e i}|^2 \epsilon_\nu \sqrt{\epsilon_\nu^2 - m^2_{i}} \Theta(\epsilon_\nu-m_i) \Bigg]. 
\label{eq:exactspec}
\end{aligned}
\end{equation}

\noindent In the electron phase space term (first bracketed term), $G_F$ is the Fermi coupling constant, $V_{ud}$ is the Cabbibo mixing angle, $M_{\rm nuc}$ is the nuclear matrix element, $E_e (p_e)$ is the outgoing electron energy (momentum), and $F(Z,p_e)$ is the Fermi function, for a daughter nucleus with charge $Z$. In the neutrino phase space term (second bracketed term), $U_{\rm e i}$ are the electron neutrino mixing matrix elements, $\epsilon_\nu$ and $\sqrt{\epsilon_\nu^2 - m^2_{i}}$ represent the total energy and momenta of the released neutrino, and $\Theta$ is the Heaviside step function.
We also define the kinetic energy of the electron, $K_e = E_e - m_e$.

In this section, we first justify our choice to hold the electron phase space term constant with respect to energy, allowing us to model spectral data by focusing on the second, neutrino-specific term. We then approximate and re-parameterize the neutrino phase space, producing an analytic spectral form that both incorporates expected features of a real data set and is suitable for Bayesian modeling.

\subsection{Approximations to the Beta Spectrum}

For this model, we consider an eV-scale energy region near the high-energy end of a spectrum produced by {$\beta$-decay}.
For tritium decay, only superallowed transitions occur, so $M_{\rm nuc}$ is simply the sum of the vector $(g_V)$ and axial vector $(g_A)$ coupling constants:
\begin{equation}
|M_{\rm nuc}|^2 = g_V^2 + 3 g_A^2.
\nonumber
\end{equation}
\noindent $M_{\rm nuc}$ is therefore independent of electron energy.

The relativistic correction to the Fermi function is negligible at these energies, so the non-relativistic form is used:

\begin{equation}
F(Z,p_e) = \frac{2\pi \alpha Z /\beta}{1-e^{-2\pi \alpha Z /\beta}},
\label{eq:FermiFunc}
\end{equation}

\noindent where $\alpha$ is the fine structure constant and $\beta \equiv p_e/E_e$ is the electron's velocity. %(The relativistic correction to $F$ is small enough to be ignored.)
Since we confine our analysis to a region of width {$\delta K_e\sim10$ eV}, and the variation in $\beta$ is of order $\delta K_e/p_e \ll p_e^{\rm max}/E_e^{\rm max}$, $\beta$ can be approximated as constant. Given Eq.~\ref{eq:FermiFunc}, then, $F(Z,p_e) \simeq F(Z,p_e^{\rm max})$.  Similarly, we treat $p_e E_e \simeq p_e^{\rm max} \cdot E_e^{\rm max}$ as constant, given that $\delta K_e \ll  E_e^{\rm max}, m_e$. Thus, we can define a constant $A \equiv \frac{G_F^2 |V_{ud}|^2}{2 \pi^3} |M_{\rm nuc}|^2 F(Z,p_e^{\rm max}) p_e^{\rm max} E_e^{\rm max}$, representing the electron phase space.  

In addition, the spectrum's neutrino-dependent term can be expressed in terms of the kinetic energy of the electron $K_e$. The neutrino phase space strongly depends on the final state of the daughter. When multiple final state configurations are possible---for example, in {\it molecular} tritium decay---all possible final state configurations need to be taken into account. In this case, however, we focus solely on {\it atomic} tritium (T) decay to singly-ionized{$~^{3}\text{He}^+$} (the process of interest for the Project 8 {experiment}~\cite{Esfahani2017}). 

Assuming the decaying source is composed of nearly pure T, we need only consider a transition to one final state configuration of the helium-3 nucleus. Energy conservation then allows us to define $\epsilon_\nu$ as
\begin{eqnarray}
\nonumber
\epsilon_\nu &\simeq& (Q_\text{T}^0 + m_e - E_{\rm recoil} - E_e) \, \equiv \, (Q_\text{T} - K_e),\\
\nonumber
Q_\text{T}^0 \, &\equiv& \, M_i - M_f - m_e -\delta b,\\
\nonumber
E_{\rm recoil}^{\rm max} &\simeq& \frac{Q_\text{T}^0 (Q_\text{T}^0+2m_e)}{2 M_f Q_\text{T}^0},
\end{eqnarray}

\noindent where $M_{i(f)}$ is the parent (daughter) nucleus mass, $\delta b$ is the difference in binding energy between the parent and daughter atoms, and $E_{\rm recoil}$ is the recoil energy of the decay nucleus (with maximum $E_{\rm recoil}^{\rm max}$). The recoil energy varies by $\sim\!0.5$\,eV over the spectrum's last 3.5 keV, so we approximate $E_{\rm recoil}$ as constant near the end of the spectrum~\cite{Bodine2015}. This allows us to write the $\beta$ spectrum in terms of an endpoint energy parameter that is assumed not to differ from decay-to-decay: $Q_\text{T} \equiv Q_\text{T}^0 - E_{\rm recoil}^{\rm max}$.  For atomic tritium, $Q_\text{T}^0$ has an experimentally determined mean value of 18566.66\,eV, and $E_{\rm recoil}^{\rm max}$ is {3.41}\,eV~\cite{Bodine2015}.

Putting this together with the constant electron phase space approximation, we formulate a spectral model ${\cal P}$: 
\begin{align}
{\cal P}(K_e) \equiv & \, A \sum_i |U_{\rm e i}|^2 (Q_\text{T} - K_e) \sqrt{(Q_T-K_e)^2-m_i^2} \nonumber
\\ & \cdot  \Theta(Q_\text{T}-K_e-m_i)  {\ \equiv \sum_i |U_{\rm e i}|^2 {\cal P}_i(K_e)}.
\label{eq:P}
\end{align}
\noindent Second-order effects are small compared with the overall spectral shape in and around our narrow analysis window. Hence, our analytic model ignores second-order corrections, including terms that account for finite nuclear radii and radiative corrections.

\subsection{\label{sec:one-two-model}One- and Two-Neutrino Spectral Models with Finite Energy Resolution}

We must transform the function ${\cal P}(K_e)$ so that it includes features seen in experimental data, including an energy resolution, background events, and { kinetic energy bounds.} In performing these transformations, ${\cal P}(K_e)$ must meet two conditions to be suitable for Bayesian inference. First, we require that the function be normalizable, because {Bayesian} models are formulated as probability density functions (PDFs). Specifically, in Stan, one specifies features of a likelihood space by adding log PDFs to a total log probability. While strictly, the function's normalization need not be analytic because Stan provides for 1D integration, inference with analytic PDFs is less computationally expensive.
By incorporating smearing from an experimental energy resolution, we are able to formulate an analytically normalized version of ${\cal P}$. Second, to assess sensitivity to the mass ordering, our model must include a parameter $\eta$, as described in Section~\ref{IIB}---or more generally, a variable that strongly depends on the ordering.

%When modeling spectral data, it is often impossible to separate the theoretical spectrum from its experimental realization.
We consider two experimental factors: the uncertainty associated with reconstructing an energy spectrum and the presence of background events. As opposed to considering an integrating spectrometer (like the one used by KATRIN), we focus on differential spectrometers (used by Project 8, ECHO and HOLMES) capable of measuring individual electron kinetic energies~\cite{Fertl2018}. This allows us to assume that true kinetic energies are normally distributed around $K$. The mapping distribution is ${\cal N}(K_e| K, \sigma)$ for a standard deviation---that is, an energy resolution---$\sigma$.

The convolution of the neutrino phase space term with ${\cal N}$ is not analytically integrable. We address this issue by expanding each neutrino mass term ${\cal P}_i$ within ${\cal P}$ (Eq.~\ref{eq:P}) to first order in $m_i^2$:
\begin{equation}
{\cal P}_i(K_e) \simeq A \cdot \big[(Q_\text{T} - K_e)^2 -  m_i^2/2\big] \Theta(Q_\text{T}-K_e-m_i).
\nonumber
\end{equation}
\noindent This expansion is justified for $m_i^2 \ll (Q_\text{T} - K_e)^2$, which holds for all data points except those very close to the endpoint.  Moreover, once the spectral shape is smeared by convolving it with ${\cal N}$, the exact and approximated curves appear very similar even near the endpoint, as seen in Figure~\ref{fig:comparespectra}. When analyzing a full spectral shape, the expansion holds except for large quantities of data. (The count number at which the approximation breaks down depends on the analysis window and binning, among other factors.)

Given the expansion in $m_ i^2$, we can define and integrate a reconstructed energy spectrum ${\cal F}_i$:

\begin{widetext}
\begin{eqnarray}
{\cal F}_i(K | Q_\text{T}, K_{\rm min}, m_i, \sigma) &\equiv& {\cal F}_i(K) \propto \int {\cal P}_i(K_e) \cdot {\cal N}(K_e| K, \sigma) \cdot \Theta(K_e - K_{\rm min}) \cdot dK_e  \rightarrow \frac{d N}{dK} \nonumber \\
   &=&  \mathfrak{N}(m_i, Q_{\rm T}-K_{\rm min}) \cdot \big[\xi(K| Q_\text{T}, m_i, \sigma, m_i) - \xi(K| Q_\text{T}, m_i, \sigma, Q_{\rm T} - K_{\rm min})\big] \label{eq:FofK} \\
\xi (K | Q_\text{T}, m_i, \sigma, t) =   (Q_\text{T}&-&K+t)\sigma^2 {\cal N}(Q_\text{T}-K|t,\sigma) + \frac{1}{2}\Bigg(-\frac{m_i^2}{2}+(Q_\text{T}-K)^2+ \sigma^2\Bigg)\cdot {\rm Erfc}\Bigg(\frac{t-Q_\text{T}+K}{\sqrt{2}\sigma}\Bigg) \nonumber
\end{eqnarray}
\end{widetext}

This model describes signal data in a kinetic energy window $[K_{\rm min}, Q_{\rm T}]$.  Its normalization term is defined based on the size {$\delta K_e$ of this window}:
\begin{equation}
\mathfrak{N}(m_i, \delta K_e) = \frac{6}{2 (\delta K_e)^3 - 3m_i^2 \delta K_e + m_i^3}
\nonumber
\end{equation}
 %\equiv Q_{\rm T}-K_{\rm min}$:

\noindent The practical need to filter out events below some energy motivates our choice to include a minimum energy parameter. Because of the {uncertainty} $\sigma$ associated with the reconstruction of $K_{\rm min}$, this lower bound is soft.

\begin{figure}[htb!]
\makebox[\linewidth][c]{\includegraphics[width=1.1\linewidth]{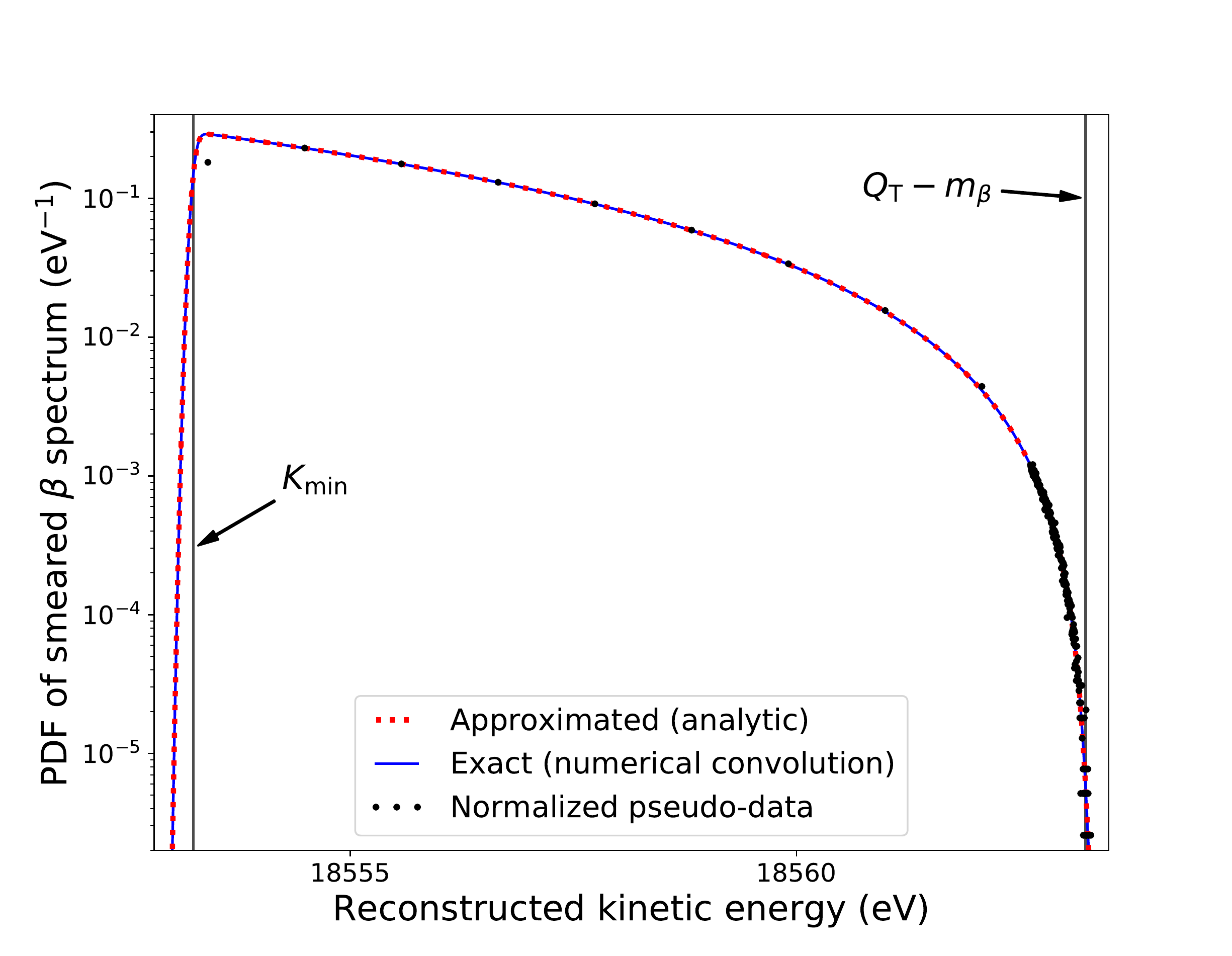}} %_ordering-mini-plot
\caption{Approximated spectral model (Eq.~\ref{eq:FofK}) superimposed on { a numerical convolution of a Gaussian with the exact T spectrum (Eq.~\ref{eq:exactspec}) and one year of data generated with the exact model.} The signal activity is $1.7\times10^8$/yr in the analysis window, $m_\beta$ = 8.5\,meV, $K_{\rm min}$ = $Q_{\rm T}\!-\!m_\beta\!-\!10$\,eV, and $\sigma$ = 54\,meV (see Section~\ref{4a}).} %The ``normal" and ``inverted" curves (numerical convolution; assumed background of $10^{-6}$/eV/s) show how a spectral shape analysis could be sensitive to the mass ordering.
\label{fig:comparespectra}
\end{figure}
% $K_{\rm min}$ = $Q_{\rm T}\!-\!m_\beta\!-\!10$\,eV,
%For the ``normal" and ``inverted" curves, a background of $10^{-6}$/eV/s is assumed.}
%Eq.~\ref{eq:smeared2nu}

The background is assumed to be uniform in kinetic energy. If we include a smeared (i.e., convolved with ${\cal N}$) background ${\cal B}$, the normalized spectral
model for a single neutrino mass $m_i$ is given by the master equation:
\begin{eqnarray}
\mathcal{M}_i(K) = f_s \cdot \mathcal{F}_i(K) + (1-f_s) \cdot \mathcal{B}(K| K_{\rm min}, K_{\rm max}, \sigma)\quad 
\label{eq:M} \\
\mathcal{B}(K| K_{\rm min}, K_{\rm max}, \sigma) = \frac{{\rm Erf}(\frac{K_{\rm max}-K}{\sqrt{2}\sigma}) - {\rm Erf}(\frac{K_{\rm min}-K}{\sqrt{2}\sigma})}{2(K_{\rm max}-K_{\rm min})}.
\nonumber
\end{eqnarray}

\noindent Here, $f_s$ is the signal fraction of a data set. Since ${\cal M}_i(K)$ is analytic and normalized, it can be formulated as a PDF and thus used for statistical inference in Stan. Moreover, Eq.~\ref{eq:M} allows us to assess an experiment's sensitivity to the mass scale. Specifically, the calibration procedure in Section~\ref{IIA} can be applied for posteriors on masses $m_i$ inferred using this model.

Experimentally, $K$ is constructed from some observed variable $v_o$---for example, in Project 8's case, an electron cyclotron frequency (see Section~\ref{IV}).  The energy resolution derives { in large part} from statistical uncertainties on the quantities used to map $v_o \rightarrow K$. While these quantities and their errors are expected to be well known, a mapping bias { that shifts the overall energy scale} is possible. We model this bias by constructing a prior on $K_\text{min}$ which allows the minimum energy to shift slightly relative to $Q_\text{T}$.

{ We bin data to reduce computation time, though un-binned analyses are possible in Stan. Details in the spectral shape on the order of a few meV only inform the neutrino mass measurement if they occur in the last $\approx\,1$\,eV. Thus, data should be binned finely near the endpoint and coarsely (for computing efficiency) at lower energies. For narrow bins, the fraction of counts per bin can be fitted to the spectral rate at each bin center. However, modeling large bins ($\mathcal{O}(1\, \text{eV})$ width) in this way biases $m_\beta$ posteriors upward relative to inputs, due to the changing slope of the spectrum within each bin. To address this, we derive the cumulative distribution function $\mathcal{G}_i^{\text{CDF}}(K)$ corresponding to the PDF model in Eq.~\ref{eq:FofK}, then set the number of events in a bin $[K_n, K_{n+1}]$ equal to $\mathcal{G}_i^{\text{CDF}}(K_n) - \mathcal{G}_i^{\text{CDF}}(K_{n+1})$. The CDF is provided in Appendix \ref{ApA}.}

To report mass ordering results based on inferred posteriors, we modify the spectral model in a second way. If one considers the smaller mass splitting ($\Delta m_{21}^2 \equiv m_2^2 - m_1^2$) to be negligible, the signal (Eq.~\ref{eq:FofK}) can be written in terms of only {\it two} neutrino masses, $m_L$ and $m_H$. Here, $m_H > m_L$, with a splitting $\Delta m^2_{ee} \equiv m_H^2 - m_L^2 \simeq |\Delta m^2_{13}| \simeq |\Delta m^2_{23}|$. The signal is then simply a weighted sum of two spectra, corresponding to the two mass scales:
\begin{eqnarray}
\mathcal{F}'(K) = \eta \cdot \mathcal{F}_L(K|Q_\text{T}, K_{\rm min}, m_L, \sigma) + \nonumber \\
(1-\eta) \cdot \mathcal{F}_H(K|Q_\text{T}, K_{\rm min}, m_H, \sigma) 
\label{eq:smeared2nu}
\end{eqnarray}
As indicated previously, $\eta$ is the fractional contribution of the lighter mass term to the spectral shape.

Since $\Delta m_{ee}^2$ is always positive, $\eta$ is the only parameter in this model that depends on the ordering. Specifically, $
\eta$ should tend toward one value ($\eta_N$) if the ordering is normal and another ($\eta_I$) if it is inverted, where
\begin{eqnarray*}
\eta_N \equiv |U_{e1}|^2 + |U_{e2}|^2 = \cos^2{(\theta_{13})}\\
\eta_I = 1-\eta_N = |U_{e3}|^2 = \sin^2{(\theta_{13})}.
\end{eqnarray*}
%~~~{\rm~for~inverted~ordering}
\noindent The ordering question can thus be formulated solely in terms of the large mass splitting
and $\theta_{13}$, both of which are measured by reactor antineutrino disappearance
experiments.  Hence, the above model enables a mass ordering determination
using only a $\beta$ spectrum and reactor experiment results. %(Such results are ordering independent, so our analysis process need not assume a mass ordering.)
%Without any external information on either $\eta$ or $\Delta m^2_{ee}$, it would be impossible to determine any ordering on the mass scale.  However, information on the measurement of $|U_{e3}|^2$ breaks this degeneracy.  Thus, by combining only the information coming mainly from reactor experiments measuring $\theta_{13}$ with a precise determination of the beta decay spectrum, it should be possible (at least in principle) to disentangle the neutrino mass ordering question simultaneously with the neutrino mass scale itself.

To perform a mass ordering sensitivity study, we substitute $\mathcal{F}_i(K) \rightarrow \mathcal{F}'(K)$ in Eq.~\ref{eq:M}. Then, by implementing the decision making scheme in Section~\ref{IIB} for posteriors on $\eta$, we can calibrate the analysis by estimating the expected accuracy of reporting different ordering results based on $\beta$ spectra. Consequently, { we have here} developed a probability distribution that serves two key purposes: It acts as a likelihood function for Bayesian modeling, and it can be used to assess a direct mass experiment's sensitivity to the mass ordering.

\section{Results}\label{IV}

%Based on Chapter 4 of my thesis and the results of an in-progress two-neutrino (ordering) analysis

Our analysis seeks to determine how experimental parameters such as energy resolution and number of $\beta$-decay events affect sensitivity to $m_\beta$ as well as the mass ordering. To construct concrete, realistic priors that reflect what parameter values an experiment might see, we incorporate information related to the Project 8 {experiment}. The Project 8 Collaboration developed the technique of Cyclotron Radiation Emission
Spectroscopy (CRES) for obtaining a $\beta$ spectrum at high precision, as originally proposed by~\cite{Formaggio2009}. CRES involves measuring the cyclotron frequencies of electrons in a magnetic field, then computing corresponding energies. In its final stage, Project 8 aims to measure the neutrino mass scale by analyzing a spectrum produced by { atomic tritium $\beta$}-decay. The Collaboration is working to reach a neutrino mass sensitivity of about 40\,meV~\cite{Esfahani2017}.
%Monreal and Formaggio

\subsection{Sensitivity to {Absolute Neutrino Mass Scale}}\label{4a}

\subsubsection{{Pseudo-Data Generation and Analysis}}

This {study} follows the procedure for calibrating sensitivity claims described in Section IIA. { We perform 220 pseudo-experiments (that is, repetitions of steps 2-5 in the procedure), assuming a runtime $\Delta t=1$\,yr. For each experiment, data is generated with a $\beta$-spectrum model that is much more detailed than the inferential model, to reveal any biases arising from analysis assumptions. The generation model includes an  energy-dependent relativistic Fermi function, as well as correction terms stemming from atomic physics phenomena. These terms account for the emitted electron's recoiling charge distribution, radiative effects from real and virtual photons, three-body recoil effects from weak-magnetism and V-A interference, 1s-orbital electron interactions with the $\beta$ and screening of the $^3$He$^+$ Coulomb field, and the $^3$He$^+$ nucleus' structure. The formulae for these corrections are taken from ~\cite{Kleesiek2019}. In this subsection, we generate data with a one-neutrino mass model and call that mass $m_\beta$.
% Kleesiek {\it et al.}, 2018 (for the KATRIN Experiment)

To compose a full data generation model, the detailed $\beta$-spectrum is broadened by numerically convolving it with a Gaussian of width $\sigma$. A nearly flat background (Eq.~\ref{eq:M}) is then added to the spectrum. Before convolution, the data is confined within a $\approx20$\,eV window centered on the mean energy at which the spectrum vanishes: $Q_\text{T} - m_\beta$, where $Q_\text{T}$ is the mean T endpoint. The window's width varies modestly from spectrum-to-spectrum because its lower bound is sampled from a prior, as discussed below.}
%(i.e., we sample a value from its prior
%$[Q_\text{T}^\mu - m_\beta - 10 \,\text{eV}, Q_\text{T}^\mu- m_\beta + 10 \,\text{eV}]$

In Stan, we implement the one-neutrino spectral model $\mathcal{M}$ from Eq.~\ref{eq:M}, for $m_i \rightarrow m_\beta$. {Each pseudo-spectrum is analyzed using this model.} 
%To achieve reasonable computation times,
The data is histogrammed with 300 bins covering the $1~\mathrm{eV}$ directly below the endpoint, nine $\approx1\,$eV-wide bins at lower energies, and one bin containing any background events above the endpoint. For each of the 300 narrow bins bounded by $[K_n, K_{n+1}]$, we model the number of counts as a value sampled from a Poisson distribution with rate $\mathcal{M}\Big(\frac{K_n+K_{n+1}}{2}\Big)\times(K_{n+1}-K_n)$. For the 9 wider signal bins, since the $\beta$-spectrum decreases monotonically, the signal Poisson rate can be approximated as $\mathcal{G^{\text{CDF}}}(K_n) - \mathcal{G^{\text{CDF}}}(K_{n+1})$ (see Appendix~\ref{ApA}). To test the effect of bin size near the endpoint, a small analysis (40 pseudo-experiments) was performed with {500 bins in the eV below the endpoint}. It yielded median $m_\beta$ sensitivities and coverages {consistent with those presented %in Table~\ref{tab:msensitivity},
below, within statistical uncertainty.}

\begin{table}
\normalsize
\begin{tabular}{ l | l | l | l}

  & Prior  & Model & Prior Source         \\
\hline
{$Q_\text{T}$} & $\mathcal{N}$([18563.25,\,0.07]eV)  & 1, 2  &   Measured     \\

{$\sigma_{\text{dopp}}$} &  $\gamma$(59.82, 2868\,eV$^{-1}$) & 1, 2   &       Measured \\

%{$\mu_{\text{inst}}$}   &  {$\gamma$(25.0, 500\,eV$^{-1}$)} & {1, 2}	     &  {Design} \\

{$\sigma_\text{inst}$}  &  {$\mathcal{N}(\mu_\text{inst}, \delta_\text{inst})$} & 1, 2	     &  Design  \\

%{$\delta_{\text{inst}}$}  &  {$\gamma$(1.583, 809.7\,eV$^{-1}$)} & {1, 2}	     &  {Design}  \\

$K_{\text{min}}$ &   $\mathcal{N}$([$Q_\text{T}\! -\! m_{\beta, L}\! {-\! 10}$,\,0.01]eV)    & 1, 2   &     Design    \\
$A_b$ & {lognorm(-27.31, 0.5678)} & 1, 2        &     Design   \\
{$N_\text{atoms}$} & {lognorm(44.07, 0.5677)}      & 1, 2   &     Design    \\
$m_\beta$	&   {$\gamma$(1.135, 2.302\,eV$^{-1}$)} & 1	    &   Measured   \\
%$m_L$	&   $\gamma$(0.838, 6.173\,eV$^{-1}$) & 2	     &  Measured  \\
$\Delta m^2_{ee}$  &   $\gamma$(314.5,\,122700\,eV$^{-2}$)  & 2		   & Measured  \\
$m_L$	&   {$\gamma$(2.186, 126.1\,eV$^{-1}$)} & 2	     &  N/A  \\
%$K_{\text{min}}$ &   $\mathcal{N}$([$Q_\text{T}^\mu-m_L-1$,\,0.01]eV)    & 2   &     Projected    \\
\end{tabular}
\caption{Priors for data generation and analysis using one- and two-neutrino models, denoted by ``1" and ``2," respectively. {``Design"} quantities reflect goals for Project 8, while ``measured" ones derive from past experiments. Prior functions are defined in Appendix~\ref{ApB}.}
\label{tab:1nupriors}
\end{table}
%for pre-generation sampling and analysis
%and the mean ($\sqrt{\text{variance}}$) of the prior on $\delta Q_\text{T}$ was taken to be the mean of the $\mu_{\delta Q}$ ($\sigma_{\delta Q}$) posterior from the final states model. 
%Phase IV will likely operate data temperature below 0.3 K, making this prior somewhat conservative.

\subsubsection{{Selection of Priors}}

Each model parameter requires an associated prior, both for sampling ``true'' values (generator inputs) and for inferring posteriors from data. By sampling from these priors repeatedly, creating an ensemble of model configurations, we can approach an analysis that accounts for the full range of possible spectra---given anticipated statistical and systematic errors. To construct priors, we {select} functional forms with boundary conditions that accord with physical limits on parameters. For positive quantities, we therefore generally chose log-normal or gamma ($\gamma$) distributions---the former when likely values span multiple orders of magnitude, and the latter otherwise. See Table~\ref{tab:1nupriors} for a summary of priors.

%    & Prior  & Model            & Prior information         \\
%\hline
%$Q_\text{T}$ & %$\mathcal{N}$([18563.25,\,0.07]eV)  & 1, 2  &    \small{Measured mean, error}     \\
%$\delta Q_\text{T}$ &  $\gamma$(59.82, 2868\,eV$^{-1}$) & 1, 2   &       \small{Mean, variance from} \\
%	    & 			&       	  		  & \small{Stan final states model}       \\

%$\sigma$  &  $\gamma$(25, 500\,eV$^{-1}$) & 1, 2	     &  \small{Projected mean, variance} \\
%$A_b$ & lognorm(-27.63, 1.40) & 1, 2        &     \small{Projected  bounds}   \\
%$A_s$ & lognorm(-5.634, 1.40)      & 1, 2   &     \small{Projected bounds}    \\
%$m_\beta$	&   $\gamma$(1.042, 1.538\,eV$^{-1}$) & 1	    &   \small{Measured bounds}   \\	
%$m_L$	&   $\gamma$(0.838, 6.173\,eV$^{\text{\scriptsize{-1}}}$) & 2	     & \small{Measured upper bound} \\
%$\Delta m^2_{ee}$  &   $\gamma$(315.5,\,122700\,eV$^{\text{\scriptsize{-2}}}$)  & 2		   & \small{Measured bounds}  \\
%All ``bounds" act as soft limits.
The one-neutrino model includes parameters $m_\beta$, $Q_\text{T}$,  $\sigma$, $K_{\text{min}}$, and $f_s$. A $\gamma$ prior on $m_\beta$ was constructed so that 1\% of its probability mass would fall below 0.008 eV, reflecting the lower bound from mass splitting measurements~\cite{PDG2020}. %used to be 2016
(This bound is not strict because of small uncertainties on those measurements.) {Ten percent of the prior mass on $m_\beta$ falls above 1.1\,eV, the 90\% confidence upper bound reported by KATRIN in 2019~\cite{KATRIN2019}.} %determined by the Mainz and Troitsk experiments~\cite{Kraus2005, Aseev2011}.

%That distribution centers $\approx 1$ standard deviation above Project 8's target resolution, to be slightly conservative---given the challenge of achieving the high magnetic field homogeneity required to reach $\sigma=0.04$\,eV. (Until now, the necessary energy resolution was assumed to be equal to the mass sensitivity.)
%$\sigma=0.04$\,eV
%The mean of the $\mu_{\delta Q}$ ($\sigma_{\delta Q}$) posterior was used as the mean ($\sqrt{\text{variance}}$) of the prior on $\delta Q_\text{T}$. 

%The endpoint was modeled as a normal distribution with mean $Q_\text{T}^\mu$ and standard deviation $\delta Q_\text{T}.
{We employ} a normal prior on {$Q_\text{T}$} but define the parameter as positive in Stan, truncating a negligible negative portion of the normal distribution. The mean of the  prior is the extrapolated tritium endpoint minus the electron mass, as calculated by Bodine {\it et al.}~\cite{Bodine2015}. The largest contribution to the $Q_\text{T}$ uncertainty is from the  T-$^3$He mass difference, which has been measured in Penning traps~\cite{Myers2015}.
%Myers {\it et al.} measured the { T-$^3$He mass difference error, or equivalently, the largest contribution to the $Q_\text{T}$} uncertainty.
That quantity serves as the {$Q_\text{T}$} prior standard deviation. 

{ We consider two energy resolution effects, summed in quadrature to yield the total resolution $\sigma$: 1) Doppler broadening $\sigma_\text{dopp}$ from translational motion of tritium atoms, and 2) Instrumental broadening $\sigma_\text{inst}$ from the process of reconstructing kinetic energies. To select a $\gamma$ prior on $\sigma_\text{dopp}$, we devised a Stan model that extracts posteriors for the mean expected energy spread due to thermal broadening ($\mu_{\text{dopp}}$) and the uncertainty on that spread ($\delta_{\text{dopp}}$), using the formulae in~\cite{Bodine2015}. We set the mean ($\sqrt{\text{variance}}$) of the $\sigma_\text{dopp}$ prior equal to the mean of a Gaussian fit to the $\mu_{\text{dopp}}$ ($\delta_{\text{dopp}}$) posterior, inferred for a $0.3000\pm0.0015$ Kelvin gas with negligible $\text{T}_2$ contamination.}

{The primary two expected contributions to the instrumental resolution are A) a cyclotron frequency measurement error and B) an uncertainty on the field value in the frequency to energy conversion. We construct a $\sigma_\text{inst}$ prior assuming that the field error $\Delta B/B\sim10^{-7}$ is the larger contribution~\cite{Esfahani2017}. In this case, $\sigma_\text{inst}\sim0.05$\,eV.
%To construct a $\sigma_\text{inst}$ prior, we consider that the dominant contribution to Project 8's instrumental resolution is expected to be an uncertainty of $\Delta B/B\sim10^{-7}$ on the field value in the cyclotron frequency to energy conversion.
As Project 8 is considering multiple energy calibration schemes, the uncertainty on $\sigma_\text{inst}$ could reasonably fall anywhere in the large range of $\approx0.5-10$\%. Accordingly, the $\sigma_\text{inst}$ prior's ``true" mean and standard deviation ($\mu_\text{inst}$, $\delta_\text{inst}$) are sampled from distributions before data generation, then fixed to their sampled values during inference. The $\sigma_\text{inst}$ prior is then $\mathcal{N}(\mu_\text{inst}, \delta_\text{inst})$. The $\mu_\text{inst}$ distribution for pre-generation sampling is $\gamma(25.0, 2\times10^{-3}\text{eV}^{-1})$, with mean 0.05\,eV and $\sqrt{\text{variance}}\!=\!0.01$\,eV. The $\delta_\text{inst}$ distribution is $\gamma(1.583, 809.7\,\text{eV}^{-1}$), selected so that 5\% of its probability mass would fall below (above) $2.5\times10^{-4}$\,eV ($5\times10^{-3}$\,eV). Combining the two sources of broadening, the mean $\sigma$ is 0.054\,eV.}
%For $\sigma_\text{inst}$,} a $\gamma$ prior with mean $0.05$\,eV and variance (0.01\,eV)$^2$ { is employed}. This is consistent with { the instrumental} resolution that Project 8 would expect if the dominant contribution were an uncertainty on its magnetic field measurement of $\Delta B/B\sim10^{-7}$~\cite{Esfahani2017}.

Experimenters can select $K_{\text{min}}$ before analysis by filtering out events above some {cyclotron frequency. %value of the observed variable $v_o$.
If the conversion ($\sigma$) to $K$ were known exactly, %and introduces {no bias,}
$K_\text{min}$ could be fixed during inference at a value computed from that frequency.}
%the pre-determined bound on $v_o$.
 {Instead, to allow for a  systematic shift in $K$ on the order of 0.01$\,$eV, we employ a normal prior on $K_\text{min}$
with that standard deviation}. %{FIX THIS EXPLANATION!} Since knowledge of Project 8's expected energy resolution will improve overtime, this is likely a conservative (large) estimate of the conversion bias; still, the prior assumes no overall bias in the energy scale calibration. 
%Because detailed experimental knowledge is needed to predict that error, to be conservative, we assume it is of the same magnitude as $\sigma$. 

We also incorporated priors associated with the spectral signal fraction. While external information does not directly inform a prior on $f_s$, it pertains more directly to signal and background activities $A_s$ and $A_b$. Here, $A_s$ ($A_b$) is the number of events per second generated by $\mathcal{F}(K)$ ($\mathcal{B}(K)$) in the window $[K_{\rm min}, Q_\text{T}]$ ($[K_{\rm min}, K_{\rm max}]$). We thus model the signal fraction as $f_s = S/(S+B)$, where $S = \Delta t \cdot A_s$ and $B = \Delta t \cdot A_b$ are signal and background Poisson event rates.

To inform the prior on $A_s$, a possible expected signal activity {in the unconvolved spectrum's last electronvolt} can be expressed in terms of both experiment-specific quantities (atomic source density $n$; effective source volume $V_\text{eff}$) and physical parameters (T half-life $\tau_{1/2}$; fraction { $f_\text{eV}$ of counts between $Q_\text{T}-m_\beta-1\,$eV and $Q_\text{T}-m_\beta$ for $\sigma\rightarrow0$}).
%branching ratio $r_{gs}$ to the daughter ground state
Following the approach in~\cite{Doe2013},
\begin{equation}
{A_s \text{ in the last eV} = n} \cdot V_\text{eff} \cdot \frac{\text{ln}(2)}{\tau_{1/2}} \cdot { f_\text{eV}}.
\end{equation}
\noindent Here, { the fraction of counts $f_\text{eV}$} in the last eV takes into account that all events observed in the last electronvolt are produced by decays to the $^3$He$^+$ electronic ground state~\cite{Bodine2015}, which comprise 70.06\% of the total tritium decay width~\cite{Williams1983}. { The detailed spectral model we developed for data generation enabled a new, precise calculation of $f_{\text{eV}}$, a quantity that has historically been central to projecting the activities of tritium-based neutrino mass experiments~\cite{KATRIN2004, Doe2013}. Assuming $m_\beta=0$, we find $f_{\text{eV}}=1.69\times 10^{-13}$ for T$_2$ and $2.06\times 10^{-13}$ for T. %The fraction increases with the neutrino mass by a factor of $\sim(1+(m_\beta/{\rm eV})^2)$.}

 For a number density $n = 10^{18}\, \text{atoms/m}^3$ and $V_\text{eff} = 10\, \text{m}^3$, target values for an experimental design scenario considered by the Project 8 Collaboration~\cite{Esfahani2017}, {the experiment would detect $\approx1.2\times10^5$ events per year above $Q_\text{T}-m_\beta-1\,$eV, and a factor of 1000 more above $Q_\text{T}-m_\beta-10\,$eV. We employed a log-normal prior on $N_\text{atoms}\equiv n \cdot V_\text{eff}$ for this scenario, setting its mode and standard deviation equal to $10^{19}$\,atoms. For a given apparatus, this allows for some variation in source density and detection efficiency. $A_s$ was then computed from $N_\text{atoms}$.}
 %The lognormal $A_s$ prior centers around $\mu_{A_s}$, with 90\% of its probability mass contained within one order of magnitude of that value. In other words, the distribution's ``soft bounds" differ by two orders of magnitude, reflecting the large uncertainties on $V_{\text{eff}}$ and $\rho$ that remain until a high-precision $\beta$ spectrum apparatus is built. %{NOTE THAT $N_\text{particles}$ PRIOR IS USED FOR GENERATION.}

The $A_b$ prior is informed by the  Project 8 Collaboration's goal for its dominant source of background to be cosmic rays passing through the tritium gas. Since the expected cosmic ray activity is approximately $10^{-12}$/eV/s for the $n$ and $V_\text{eff}$ values assumed above, and the activity varies with those parameters~\cite{Esfahani2017}, the $A_b$ prior {distribution is chosen to have mode and standard deviation equal to $10^{-12}$/s for each 1-eV-wide bin of data.} %so that 90\% of its probability mass falls between $10^{-13}$/s and $10^{-11}$/s
% Note that we devised a Stan model to simulate the endpoint distribution \& extract a posterior on $\delta Q_\text{T}$

%{Numbers to be updated throughout the remainder of this subsection.}
%Each pseudo-data set strongly informed a posterior on $m_\beta$; in other words, the $\beta$-spectrum provided useful neutrino mass information. As is evident from 

\subsubsection{{Neutrino Mass Scale Sensitivity Results}}

{A close correspondence between ``true" neutrino masses and $m_\beta$ posteriors indicates that each $\beta$-spectrum strongly informs a neutrino mass determination (see Figure~\ref{fig:mbeta}). Each} posterior standard deviation on $m_\beta$ is at least {22} times smaller than the corresponding prior spread. See~\cite{Betancourt2018, Gelman2013} for more information on posterior shrinkage and evaluating model performance.

%\begin{table}[htb!]
%\begin{center}
%\begin{tabular}{| r || c | c |}
%\hline
%			   & Sensitivity (eV) & Coverage
%\tabularnewline
%\hline
%\hline
%Standard dev. & 0.0109 & N/A  \\
%90\% C.I. & 0.0344 & 52\% \\
%95\% C.I. & 0.0407 & 70\%  \\
%$Q^{\text{in}}_\sigma=$1 eV & 33.6 | 39.9 	&  13.7 | 17.7 \\
%\hline
%\end{tabular}
%\end{center}
%\vspace{-0.5cm}
%\caption{{Table to be updated.} Sensitivity to $m_\beta$ after one year, with coverages of 90 and 95\% credible intervals.}
%\label{tab:msensitivity}
%\end{table}

{Table~\ref{tab:msensitivity} summarizes credible interval width results for $m_\beta$. Highest density credible intervals (C.I.s) were computed for $\alpha=0.6826$, 0.9 and 0.95 (see Eq.~\ref{eq:LossMbeta}), and standard deviations were computed by halving the first of these. The HDI approach produces higher coverages than do quantile intervals. To enable reliable C.I. estimation,} we required the effective size of each posterior array (as computed by PyStan~\cite{Stanual2020}) to exceed 6000, so that at least 150 effective samples fall outside each bound.

{We can verify that the process of inference itself was successful: As expected, posterior means for $Q_\text{T}$, $\sigma_\text{inst}$, $\sigma_\text{dopp}$, $K_\text{min}$, $A_s$ and $A_b$ track with input values. During all 220 analyses, the five Stan convergence diagnostics---$\hat{R}$, effective sample size ratio, E-BFMI, tree depth, and divergences~\cite{PystanWorkflow, Betancourt2015, MCMCinPractice}---showed no signs of pathological behavior.
%and posterior credible intervals for each parameter generally include those inputs.
Moreover, the coverage of 90\% credible intervals is between 85\% and 99\% for all parameters.}

%~\footnote{Average and median energy windows differ by $\leq$\,0.0083 eV}
%likely due to an exponential rise in the number of events between $Q_\text{T}-m_\beta$ and $Q_\text{T}$

For true $m_\beta>0.5$\,eV, the mean 90\% C.I. width is 0.005\,eV. The reported coverage uncertainties are $\sqrt{C\cdot(1-C)/N_\text{trial}}$.}

%This sensitivity analysis suggests that Project's will have a neutrino mass sensitivity near that expected from analytical predictions~\cite{Doe2013,Esfahani2017}.
%This analysis suggests that a direct mass experiment can measure neutrino masses below its energy resolution, {since the $m_\beta$ posterior standard deviation is at most 0.0158\,eV, while the inputted $\sigma_\text{inst}$ ranges from 0.02 to 0.09\,eV.} {Include note about operating in a statistics vs.~systematics dominated regime.}
%This is likely explained by our model's ability to account for the spectrum's full shape, as opposed to a shift in endpoint energy, alone. Still, an $m_\beta$ result with 95\% credibility is likely to be inaccurate 30\% of the time. This coverage could potentially be improved as new information enables stronger priors to be constructed. In particular, a 13 run analysis with a widened $Q_\text{T}^\mu$ prior of $\mathcal{N}$(18563.25\,eV, 1\,eV) yielded a coverage of only 30.8\% [90\% C.I.], indicating that better knowledge of $Q_\text{T}^\mu$ might boost the accuracy of $m_\beta$ results.
%In addition, $m_\beta$ coverage could be improved by refining the process by which initialization parameters are chosen$\scalebox{1.75}[1.0]{\( - \)}$specifically, by using a spectrum to initialize each analysis run in the appropriate region of parameter space.

The left plot of Figure~\ref{fig:msensitivity} shows that mass sensitivity depends weakly on $\sigma_\text{inst}$, because the scenario considered here is relatively statistics-limited and the range in $\sigma_\text{inst}$ is small. However, for this scenario, smaller uncertainties on $\sigma_\text{inst}$ noticeably improve sensitivity (see Section~\ref{nonzero-mass} for an instance of this). We would also expect increasing the effective volume to improve neutrino mass sensitivity. Indeed, for an ensemble with fixed energy resolution and a wide range in $V_\text{eff}$ values, the widths of $m_\beta$ credible intervals depend strongly on $V_\text{eff}$, as seen in the right plot in Figure~\ref{fig:msensitivity}.  These results could inform how future direct mass experiments prioritize their efforts to improve the expected energy resolution, resolution uncertainty, and statistical yield of an apparatus design.
%with smaller energy resolutions allowing for more precise mass determinations.

\begin{figure}[tb!]
\includegraphics[width=\linewidth]{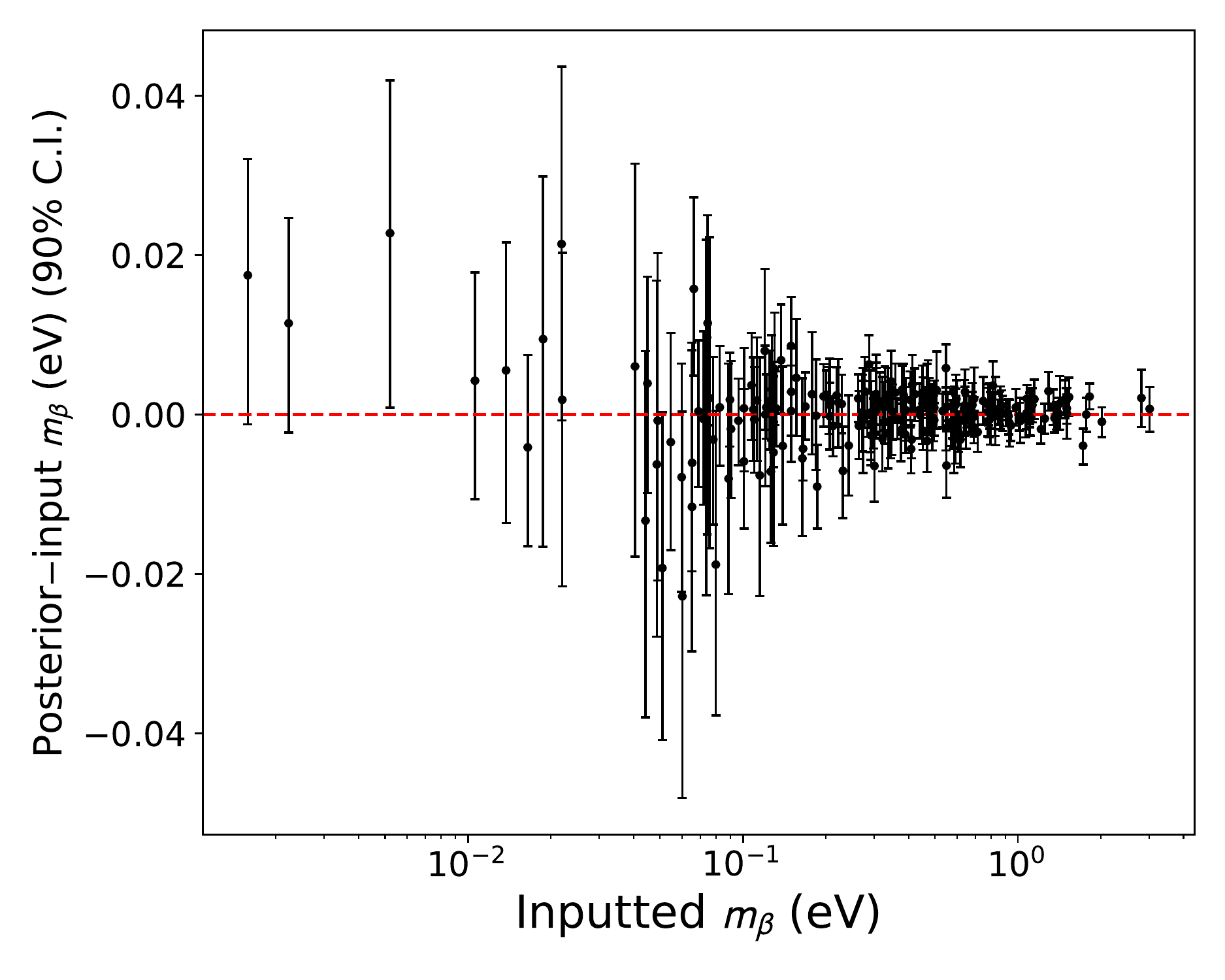}
\caption{Neutrino mass posterior means and {90\%} credible intervals as a function of inputted $m_{\beta}$, for a one-neutrino model and the assumed experimental design. Interval widths (``sensitivities") decrease with $m_{\beta}$, asymptoting at $\sim\!5$\,meV.}
\label{fig:mbeta}
\end{figure}

\begin{figure*}
\subfloat{\includegraphics[width=.473\linewidth]{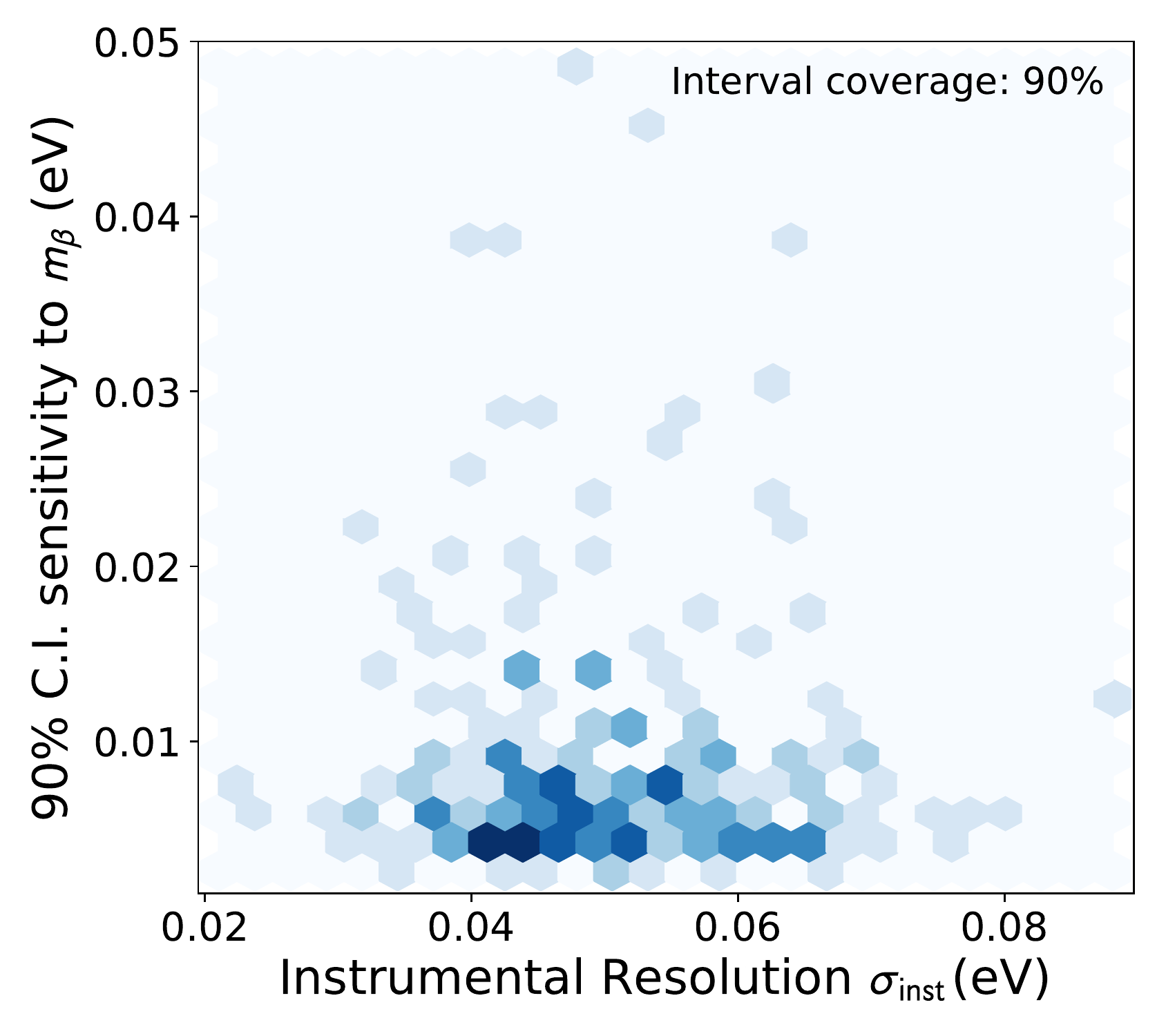}}
\subfloat{\includegraphics[width=.527\linewidth]{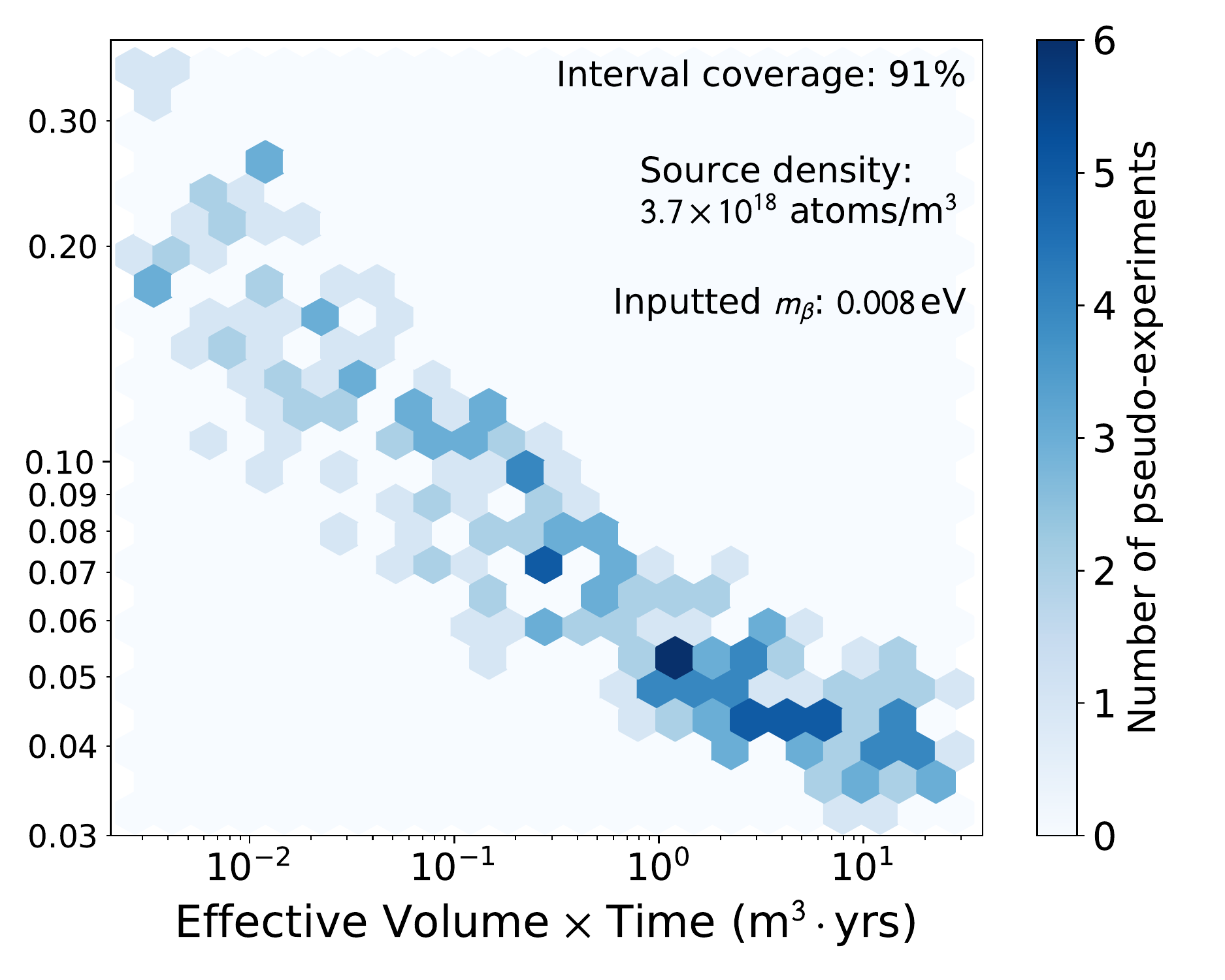}} %cenpa_90-mass-sensitivity-vs-Nparticles-hexagons.pdf
\caption{Dependence of mass sensitivity ({width of 90\% credible intervals}) on {$\sigma_\text{inst}$ and volume$\times$efficiency$\times$time}. {The left plot assumes the design scenario described in this section. The right plot shows a larger range in signal exposure, for an alternate scenario where $n$ ($3.7\times10^{18}$\,m$^{-3}$) and $\sigma$ ($115\pm2$\,meV) are chosen to minimize $m_\beta$ uncertainty, given a trade-off between frequency reconstruction error and exposure. The right plot ``pessimistically" assumes $m_\beta=0.008$\,meV.}}
%This yields 91\% coverage for 90\% C.I.s on $m_\beta$.
%and the experimental design scenario assumed here. {For a higher density source, the x-values in the right hand plot could decease without affecting sensitivity.}}

\label{fig:msensitivity}
\end{figure*}

\subsubsection{Claiming $m_\beta$ is Inconsistent With Zero}\label{nonzero-mass}

{ We also evaluate the ability of an experiment with the design described here to distinguish the electron-weighted neutrino mass from zero. As introduced in~\ref{IIA}, for a given $\beta$-spectrum, it is possible to claim that the neutrino mass is nonzero with credibility $\alpha$ if the lower bound of a posterior highest density $\alpha$-credible interval exceeds zero. The $m_\beta$ prior in Table~\ref{tab:1nupriors} is in conflict with this test, as that prior assumes that it is highly improbable for the mass to be zero, considering the lower bound from oscillations measurements. When Project 8 analyzes real data, its main mass scale analysis can include an $m_\beta$ prior with an oscillations-based lower bound. However, to assess consistency with zero, the data will need to be re-analyzed with an oscillations-bound-free prior.

\begin{table}[tb!]
\begin{center}
\begin{tabular}{| l || l | l |}
\hline
\textbf{Interval} & \textbf{Sensitivity (eV)} & \textbf{Coverage}
\tabularnewline
\hline
\hline
			 & \underline{Median} \quad  \underline{Mean} \quad \  \underline{Maximum} &  \\
90\% C.I. & 0.0071 \quad \  0.0112  \quad \ 0.0493 & $(90.0\pm2.0)$\% \\
95\% C.I. & 0.0084 \quad \  0.0133 \quad \ 0.0598 & $(93.2\pm1.7)$\%  \\
Stdev. & 0.0022 \quad \ 0.0034 \quad \ 0.0158 & $(70.1\pm3.1)$\%  \\
%$Q^{\text{in}}_\sigma=$1 eV & 33.6 | 39.9 	&  13.7 | 17.7 \\
\hline
\end{tabular}
\end{center}
\vspace{-0.5cm}
\caption{Sensitivity to $m_\beta$ after 1\,yr, with coverages {of credible intervals.}}
\label{tab:msensitivity}
\end{table} 
%Alternatively, $m_\beta$ can be resolved within a 0.041 eV window (95\% C.I.) with 70\% coverage. 
%$n_\text{eff}$

As an example sensitivity study, we perform 75 pseudo-experiments with 10\% of the neutrino mass prior probability falling below 0.005\,eV and 10\% above 0.1\,eV. Resulting posterior credible intervals on $m_\beta$ are shown in Figure~\ref{fig:checkifnonzero}. The neutrino mass can be distinguished from zero with 90\% credibility in 65 of these analyses. It is possible to claim the mass is inconsistent from zero for true $m_\beta\gtrsim0.04$\,eV, with two outliers caused by an underestimation of the true mass, combined with poor $m_\beta$ precision due to large inputted uncertainties (i.e., prior widths) on $\sigma_\text{inst}$.}
%These outliers from the tail sample for the uncertainty on $\sigma_\text{inst}$ could be moderated with a sufficiently large ensemble of pseudo-experiments.

{How can one be confident that this method will not produce frequent false claims? We may perform another calibration: For $\beta$-spectra produced given a true neutrino mass of zero, we should rarely claim that $m_\beta$ is distinguishable from zero. Indeed, when we analyze 150 such spectra, the mass is judged to be consistent with zero 93\% of the time ($\alpha=0.9$).}

\begin{figure}[htb!]
\includegraphics[width=\linewidth]{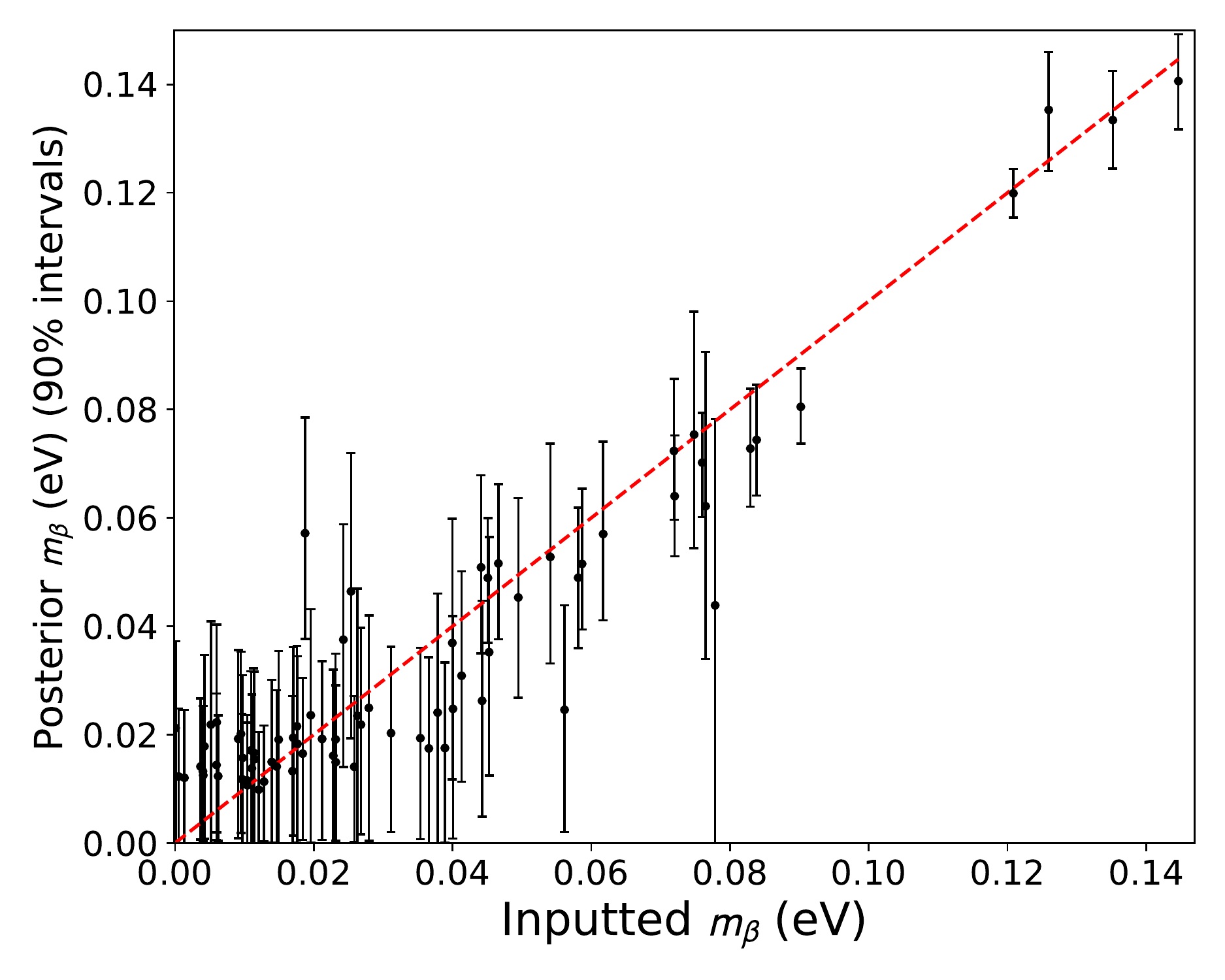}
\caption{{Mass posterior means and {90\%} credible intervals for inputted $m_\beta$ near zero. It is possible to distinguish the mass from zero for true $m_\beta\!\gtrsim\!0.04$\,eV, with outliers characterized by large uncertainties $\delta_{\text{inst}}$ (energy broadening standard dev.).}}
\label{fig:checkifnonzero}
\end{figure}

\begin{figure*}[t!]
%\makebox[\linewidth][c]{\includegraphics[width=1.15\linewidth]{2nu-model_2D-histograms.pdf}}
%\includegraphics[width=\linewidth]{2nu-model_2D-histograms_5-10A_28.png} %Another option: 2nu-model_2D-histograms_5-7_17.png
\includegraphics[width=\linewidth,trim=5 5 5 5,clip]{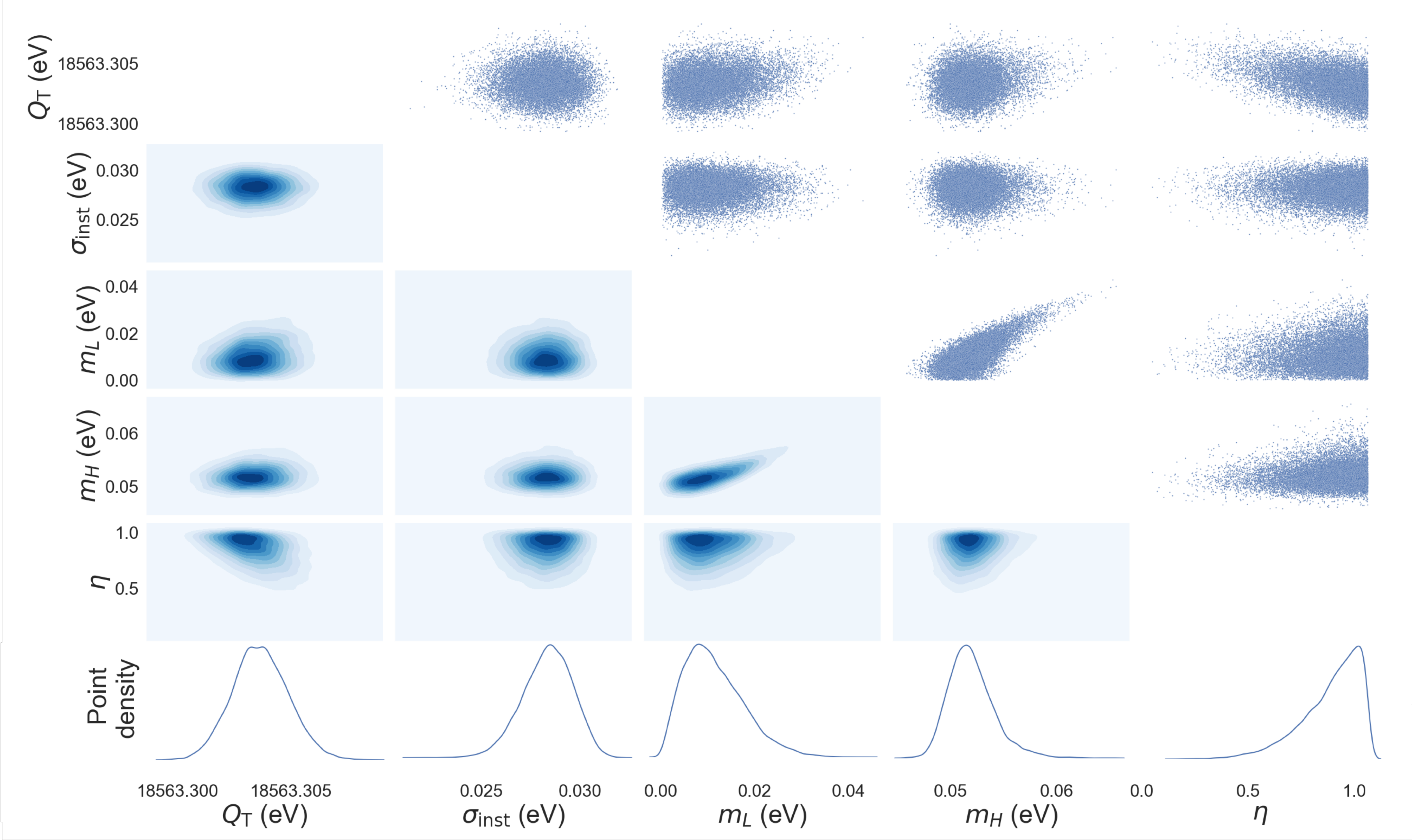} % modified version following Martin's comment

\caption{For one pseudo-experiment, example posterior probability density plots and 2D-histograms (in both contour and scatter plot form) for parameters in the two-neutrino spectral model. Posteriors were obtained by analyzing data {($\Delta t =1$\,yr)} that was generated assuming a normal mass ordering.}
\label{fig:posteriors}
\end{figure*}

\begin{figure}[htb!]
\includegraphics[width=\linewidth]{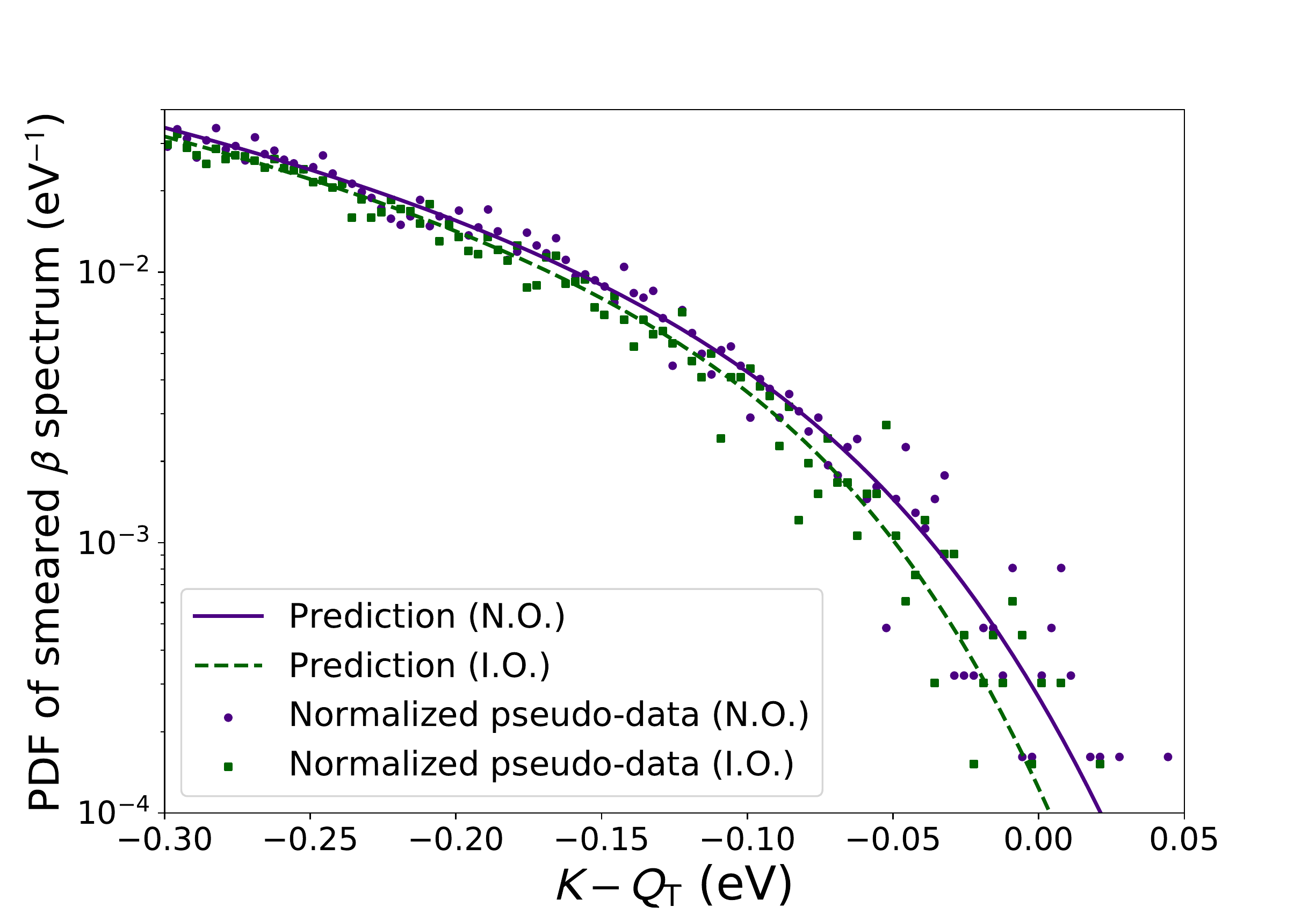} 
\caption{Example pseudo-spectra overlaid on predicted curves (Eq.~\ref{eq:exactspec} numerically convolved with a Gaussian) for normal and inverted orderings, with $m_L=0~\mathrm{eV}$ and a 2 yr runtime. Spectra are plotted as a function of the difference between reconstructed energy and the T endpoint.}
\label{fig:MO-data}
\end{figure}

\subsection{Sensitivity to Neutrino Mass Ordering}\label{sec:ordering-results}

The analysis in this section follows the procedure described in Section IIB for calibrating sensitivity claims to discrete parameters. {Pseudo-data is generated with the same detailed spectral model as in Section~\ref{4a}, but with two neutrino masses instead of one. Similarly, for inference in Stan,} we now employ a two-neutrino model---Eq.~\ref{eq:M}, with a spectral signal $\mathcal{F}'(K)$ (Eq.~\ref{eq:smeared2nu})---to analyze data in the approximate window $[Q_\text{T} - m_L - 1 \,\text{eV}, Q_\text{T} - m_L + {10} \,\text{eV}]$. This region {extends only 1\,eV below the endpoint so that the likelihood will be strongly informed by fine-grained mass ordering-dependent features near $Q_\text{T}$. To help constrain the overall mass scale, data in the next eV below $Q_\text{T} - m_L - 1 \,\text{eV}$ are fitted to a one-neutrino $m_\beta$ model, with the requirement $m_\beta^2=\eta \cdot m_L^2 + (1-\eta)\cdot m_H^2$.}

We repeat this two-neutrino analysis for $\Delta t=$\,1 yr and 2 yrs with at least {170} pseudo-experiments per runtime, producing coverage uncertainties of 1-5\%. Again, data are binned after generation, then analyzed assuming { Poisson-distributed events. The Stan} model includes the same priors on parameters {$Q_\text{T}$, $\sigma_\text{dopp}$, $\sigma_\text{inst}$, $N_\text{atoms}$} and $A_b$ as in the one-neutrino case. The prior on $K_{\text{min}}$ is similar, with its mean dependent on $m_L$ instead of $m_\beta$. We also constructed priors on $\Delta m^2_{ee}$ and $m_L$ (see Table~\ref{tab:1nupriors}), while $m_H$ required no prior, as it was modeled by transforming those parameters.\footnote{To avoid non-invertible transforms and the need for Jacobian adjustments, in Stan, we define a ``\texttt{positive\_ordered}'' \texttt{transformed parameter m}, with \texttt{m[1]}=$m_L$ and \texttt{m[2]}=$\sqrt{m_L^2 + \Delta m^2_{ee}}$ (see Section 22 of ~\cite{Stanual2020}).
The entries of 
\texttt{m} then serve as inputs to the spectral log probability density function.}

A $\gamma$ prior on $\Delta m^2_{ee}$ was formulated by extracting a 90\% confidence interval from a global fit of three reactor neutrino experiments: $[2.38,2.75]\times10^{-3}\,\text{eV}^2$~\cite{GlobalFit}. At the time when we began the analysis, this was the most up-to-date global fit of reactor data. As these bounds differ slightly according to mass ordering, to be conservative, we selected each bound (either the normal or inverted ordering limit) so as to obtain a wider prior. Ten percent of the prior mass on $\Delta m^2_{ee}$ falls outside each bound. In addition, before generation, either $\eta_N$ or $\eta_I$ was sampled from a Gaussian prior, depending on the ``true" ordering. Prior parameters were determined based on the mean of $\cos^2\theta_{13}$ (0.979) and error on that mixing parameter (0.001), as measured by reactor experiments~\cite{GlobalFit}.  Posteriors extracted from one of the two-neutrino model fits are shown in Fig.~\ref{fig:posteriors}, and Fig.~\ref{fig:MO-data} compares pseudo-datasets for the normal and inverted orderings.

For the prior on $m_L$, we avoided computing soft bounds using current limits on the mass scale from particle physics experiments, as those constraints do not translate easily to bounds on individual masses~\cite{PDG2020, Abazajian2011}. % PDG2018, Abazajian2011
Instead, we envision a scenario in which $m_L$ is restricted below $\approx 0.05\,$eV, potentially based on future cosmological constraints on the sum of the three neutrino masses. Specifically, the prior for pre-generation sampling and inference is $\gamma$-shaped with 10\% of its mass below 5\,meV and {5\% above 40\,meV, resulting in $m_L < 0.08$\,eV} for all pseudo-experiments. (The distribution peaks near zero, since there is no oscillations-based lower bound on $m_L$ for the inverted ordering and a very small lower bound for the normal case.) {For true masses above $0.08$\,eV, one rarely claims} to have resolved the mass ordering using our reporting scheme. Hence, by choosing a prior localized in a low-mass region, we proportionally inflate true and false ordering claim rates. This makes the process of selecting {ideal reporting criteria} $\kappa$ based on claim rates more statistically reliable than it would be for a wider $m_L$ prior.

%For comparison, we repeat our analysis with such a wider prior, for $\Delta t=$ 1 year. In this case, we make use of the maximum upper limit typically obtained from cosmological data: 1.3\,eV~\cite{PDG2018}. Assuming mass splittings are negligible at this scale, the soft bound on $m_L$ would be $\frac{1.3}{3}\approx 0.43$\,eV. Thus, the wide mass prior was constructed so that 5\% of its mass fell above 0.43\,eV and 10\% below 0.01\,eV.
%Instead, the soft upper limit on $m_L$ is taken to be one-third of the bound on the sum of the three neutrino masses from cosmological data. This is reasonable assuming mass splittings are negligible relative to overall scale, which is the case near the maximum bound typically obtained: $\approx$\,1.3\,eV~\cite{PDG2018}. Thus, the prior was constructed so that 5\% of its mass fell above $\frac{1.3}{3}\approx 0.43$\,eV. In addition, 10\% of its mass fell below 0.01\,eV, so that the distribution peaks near zero (the only lower bound on $m_L$).

 {As in the one-neutrino case, posterior means track with input values for all parameters. During analysis of most spectra, no Stan MCMC diagnostics indicated a failure to converge. However, a quarter of runs exhibited signs of incomplete convergence~\cite{PystanWorkflow}: 15\% showed a small number of diverging iterations (1-10 of 15,000), and 10\% failed at least one other check. Mass ordering sensitivity results are robust despite this, since observing and minimizing false positive rates ultimately validates the analysis. Still, a more consistently converging model might improve sensitivity.}
 %We therefore eliminated 1 of 130 four-years run and 3 of 136 two-year runs, during which at least 1\% of iterations exceeded this limit. This occurred for $\approx$\,1-10 of 12,000 iterations in a few other cases. In addition, during approximately a fourth of the generation runs, a small number (less than $0.001$\%) of iterations diverged.

%\begin{figure}[htb!]
%\includegraphics[width=0.9\linewidth]{9-1_eta-90-post-vs-mL.pdf}
%\caption{Posterior means and intervals on $\eta$ as a function of $m_L$, for a two-neutrino model and two years of data. For data generated assuming a normal ordering, at low $m_L$, we can rule out $\eta$ values expected for the inverted ordering (near 0).}
%\label{fig:etavmL}
%\end{figure}

\begin{figure}[htb!]
\includegraphics[width=\linewidth]{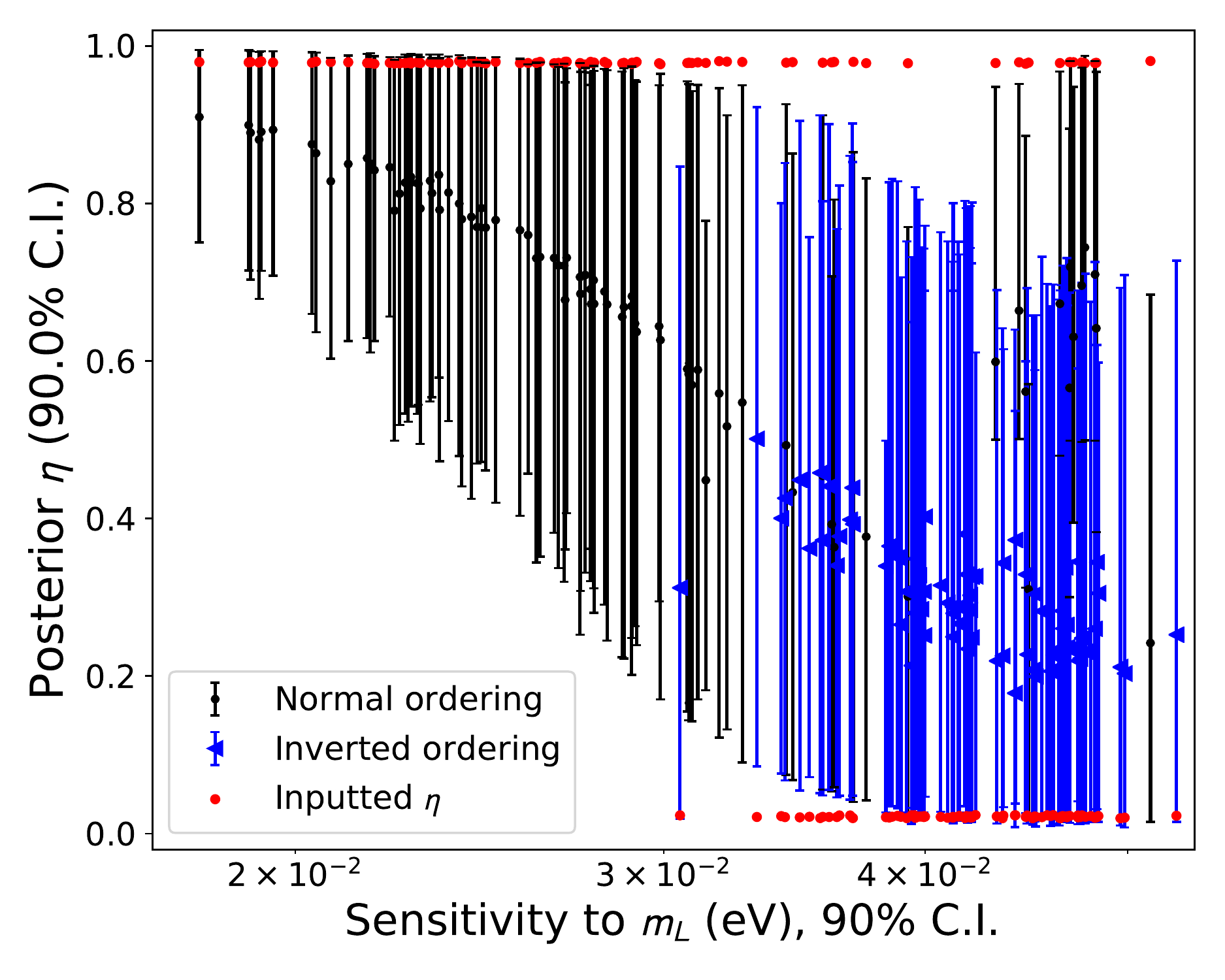} %2-24_eta-88-post-vs-msensitivity_1yr.pdf
\caption{For 2 yrs of data assuming normal (dots) and inverted (triangles) orderings, posterior means and intervals on $\eta$ as a function of potential sensitivity to $m_L$, defined as C.I. width.}
\label{fig:etavsensitivity}
\end{figure}

%including in the comparison case with a wide $m_L$ prior. 
Table~\ref{tab:mHsensitivity} summarizes results for calibration of sensitivity to the mass {ordering. Uncertainties on 0\% claim rates represent 68.3\% confidence limits derived from a binomial probability law. (Given the ensemble's finite size, the actual probability of a false claim is not exactly zero.) % by solving for $p$ in $Pr(X=0)=(1-p)^{N_{\text{trial}}}<(1-0.683)$.
The} loss functions $L_N$ and $L_I$ in Eq. 3 dictated whether an ordering result should be reported for each pseudo-experiment. That is, a normal (inverted) ordering claim was made if a posterior interval on $\eta$ of credibility $\kappa$ contained $\eta_N=\cos^2\theta_{13}$ ($\eta_I=1-\eta_N$) but not $\eta_I$ ($\eta_N$) (see Figure~\ref{fig:etavsensitivity}). Given the small experimental error on $\cos^2\theta_{13}$, we assumed a known value $\eta_N=0.978$. The credibility $\kappa$ acts as a reporting criterion, and modifying $\kappa$ affects the rates at which we {\it correctly} and {\it incorrectly} claim to have resolved the neutrino mass ordering (see Figure~\ref{fig:kappaplot}).

\begin{table}[tb!]
\begin{center}
%$\kappa=0.88$, $\Delta t = 4$\,yrs, $m_L\lesssim0.05$\,eV %86 runs
%\begin{tabular}{| r || c | c | c |}
%\hline
%			   & Claim $N$  & Claim $I$ & No Claim
%\tabularnewline
%\hline
%\hline
%Truth: $N$  & 47.3\% & 0.0\% & 53.7\%  \\
%Truth: $I$ & 0.0\% & 0.0\% & 100\% \\
%\hline
%\end{tabular}

\newcommand\Tstrut{\rule{0pt}{2.2ex}} 
\newcommand\Bstrut{\rule[-0.9ex]{0pt}{0pt}}

%\vspace{8 pt}
$\Delta t = 2$\,yrs, $m_L\lesssim0.05$\,eV

\begin{tabular}{| r ||c|c|}
\hline
			   & Claim $N$  & Claim $I$ 
\tabularnewline
\hline
\hline
Optimal $\kappa$ & 0.985 & 0.855 \Tstrut \Bstrut \\
\hline
Truth: $N$ & $86.8\%\pm3.5\%$ & \,  $0.0\%$ (+1.3\%) \Tstrut  \\
Truth: $I$ & \, $0.0\%$ (+1.5\%)\, & $21.8\%\pm4.7\%$ \\
\hline
\end{tabular}

\vspace{8 pt}
$\Delta t = 1$\,yr, $m_L\lesssim0.05$\,eV

\begin{tabular}{| r ||c|c|}
\hline
			   & Claim $N$  & Claim $I$ 
\tabularnewline
\hline
\hline
Optimal $\kappa$ & 0.925 & 0.875 \Tstrut \Bstrut \\
\hline
Truth: $N$ & $ 45.6\%\pm5.2$\% & \, $0.0\%$ (+1.3\%)  \Tstrut  \\
Truth: $I$ & \, $0.0\%$ (+1.2\%)\, & $23.5\%\pm4.3\%$  \\
\hline
\end{tabular}

%\vspace{8 pt}
%$\kappa=$, $\Delta t = 1$\,yr, $m_L\lesssim0.43$\,eV %91 runs
%\begin{tabular}{| r || c | c | c |}
%\hline
%			   & Claim $N$  & Claim $I$ & No Claim
%\tabularnewline
%\hline
%\hline
%Truth: $N$  & \% & 0.0\% & \% \\
%Truth: $I$ & \% & 0.0\% & \% \\
%\hline
%\end{tabular}
\end{center}
\vspace{-0.5cm}
\caption{Assuming either a normal or inverted true ordering, percentages of pseudo-experiments for which the three possible reporting outcomes (``normal," ``inverted," or ``no claim") occur. {To minimize false claims, different reporting criteria $\kappa$ are used for each ensemble and observed ordering.}}
\label{tab:mHsensitivity}
\end{table}

\begin{figure}[tp]
\subfloat{%
  \includegraphics[clip,width=\columnwidth]{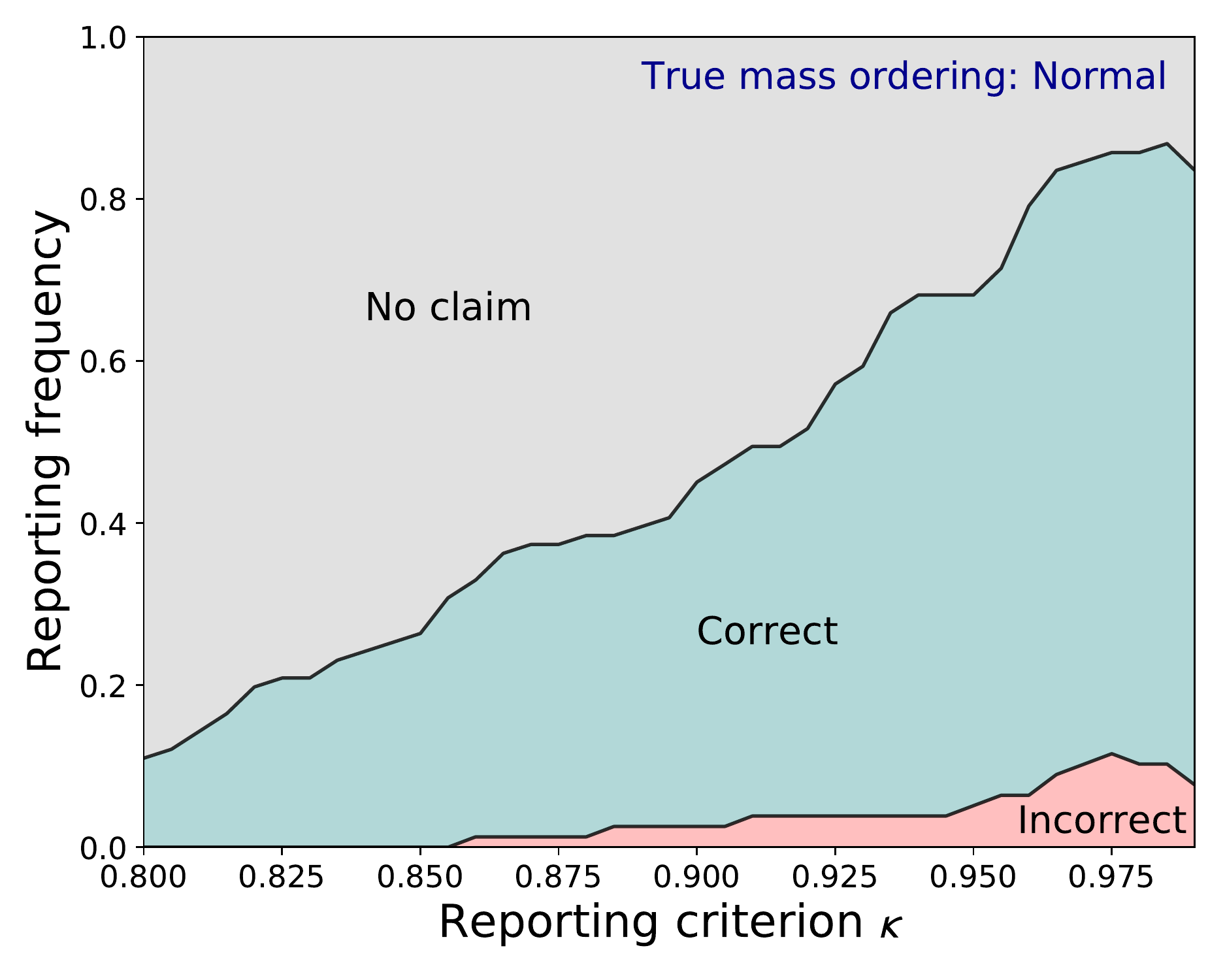}%
}

\subfloat{%
  \includegraphics[clip,width=\columnwidth]{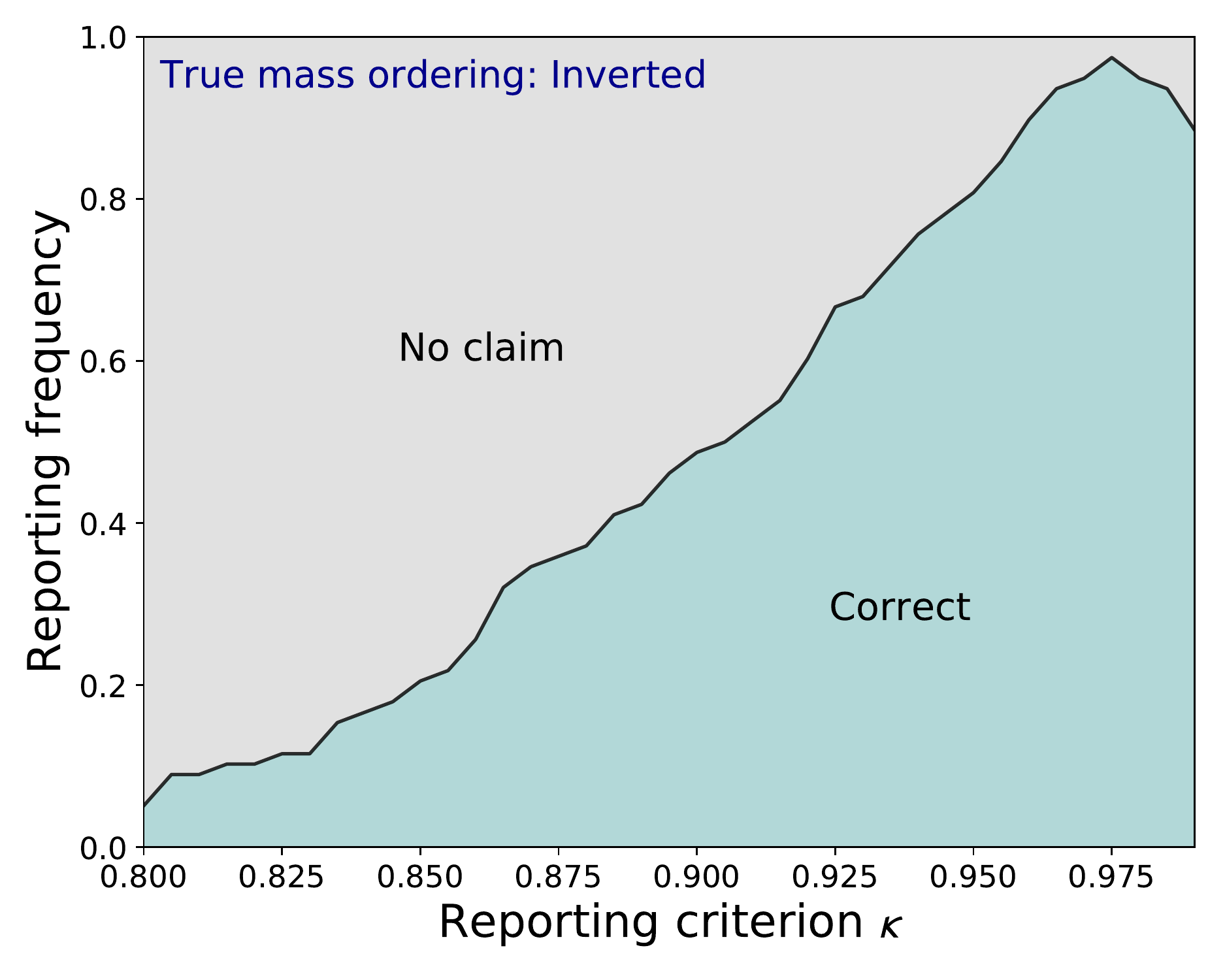}%
}
\caption{Mass ordering reporting frequencies for $\Delta t=2$\,yrs as a function of $\kappa$, the credibility of the $\eta$ interval (see Eq.~\ref{eq:Loss-ordering}). To obtain the rates in Table~\ref{tab:mHsensitivity}, different $\kappa$ values are chosen depending on whether the initially favored result is normal or inverted. For the upper plot, this adjustment enables one to reduce the incorrect claim rate.}
\label{fig:kappaplot}
\end{figure}
%that determines whether a result is claimed

%An {``optimal criterion" should be selected given an} understanding of what frequencies of correct and incorrect claims are acceptable. Accordingly,
{We recommend an ``optimal $\kappa$" by selecting the value for which the relevant correct claim rate is maximized, given a minimal incorrect rate---which can be zero, in this study. Values of $\kappa$ are considered in 0.5\% increments. We observe that, for both 1 yr and 2 yrs of data, false inverted claims begin to occur for $\kappa$ values above a {\it lower} number than do false normal claims. In fact, Figure~\ref{fig:kappaplot} shows that false normal claims are never made for $\Delta t=2\,$yrs, for these pseudo-data sets. Using that knowledge, for real data, it is possible to boost the probability of a correct ordering claim without increasing the risk of a false claim by applying the following procedure:
\vspace{-2pt}
\begin{enumerate}[label=\Alph*)]
\itemsep0em 
\item Check what result would be reported using the optimal $\kappa$ for normal ordering true/false claims (as predicted with pseudo-experiments)---here, 0.925 (0.985) for 1 (2) yr(s). 
\item If the result is ``normal" or ``no claim," report it.
\item If the result is ``inverted," it could be a false positive. Reduce $\kappa$ to the inverted optimal value---0.875 (0.855) for 1 (2) yr(s)---to determine if to report ``inverted" or nothing.
\end{enumerate}
\vspace{-4pt}
\noindent This procedure accounts for the fact that it is easier to claim a normal than an inverted ordering result for our model. The procedure enables false claim rates of 0\% for the pseudo-experiments performed here, with true rates reaching 87\% (22\%) for the normal (inverted) ordering after 2 yrs.  We see here that as statistical power improves over time, sensitivity to the normal ordering improves.}

%\begin{figure}[htb!]
%\includegraphics[width=\linewidth]{2nu-2yr_kappa_reporting.pdf} %3-17_kappa_reporting_4yrs_final.pdf
%\caption{Mass ordering reporting frequencies for $\Delta t=2$\,yrs as a function of $\kappa$, the credibility of the $\eta$ interval that determines whether a result is claimed.}
%\label{fig:gammaplot}
%\end{figure}

%$82.5\% < \kappa <83.5\%$
%For one year of data taking, this condition is met for  $\kappa =80.5\%$. Thus, if researchers on a direct mass experiment with data described by our model wished to report an ordering result after around one year, we would recommend they require that an 80.5\% credible interval on $\eta$ include exactly one of ($\eta_N$, $\eta_I$). For two (four) years of data taking, $\kappa=87.5\%  \ (88)\%$ meets our requirement, so we would advise employing these criteria for longer runtimes. As expected, we see here that as statistics improve over time, it is possible to select a less stringent (higher) $\kappa$ while maintaining a low false positive rate.
%determined by taking the mean of $\cos^2\theta_{13}$ for each ordering, as determined by reactor experiments~\cite{GlobalFit}
%In addition, we require that the total false claim rate not exceed 1\%, since otherwise the ``ideal" criterion could easily lead experimenters to report the wrong mass ordering.

%These results indicate that it is difficult to reliably report an ordering result with one year of data.
These results indicate that a Project 8-like neutrino mass experiment could resolve the mass ordering for various likely combinations of physical and experimental parameter values. If the neutrinos obey a normal ordering and the lightest mass is constrained below $\approx0.05$\,eV, this analysis predicts there is a {high chance of resolving the ordering after 2 yrs of data taking. We observe that a} direct mass experiment would resolve the normal ordering especially often in the low-$m_L$, small mass sensitivity region (see Figure~\ref{fig:MHreporting}). 

\begin{figure}[tb!]
\includegraphics[width=\linewidth]{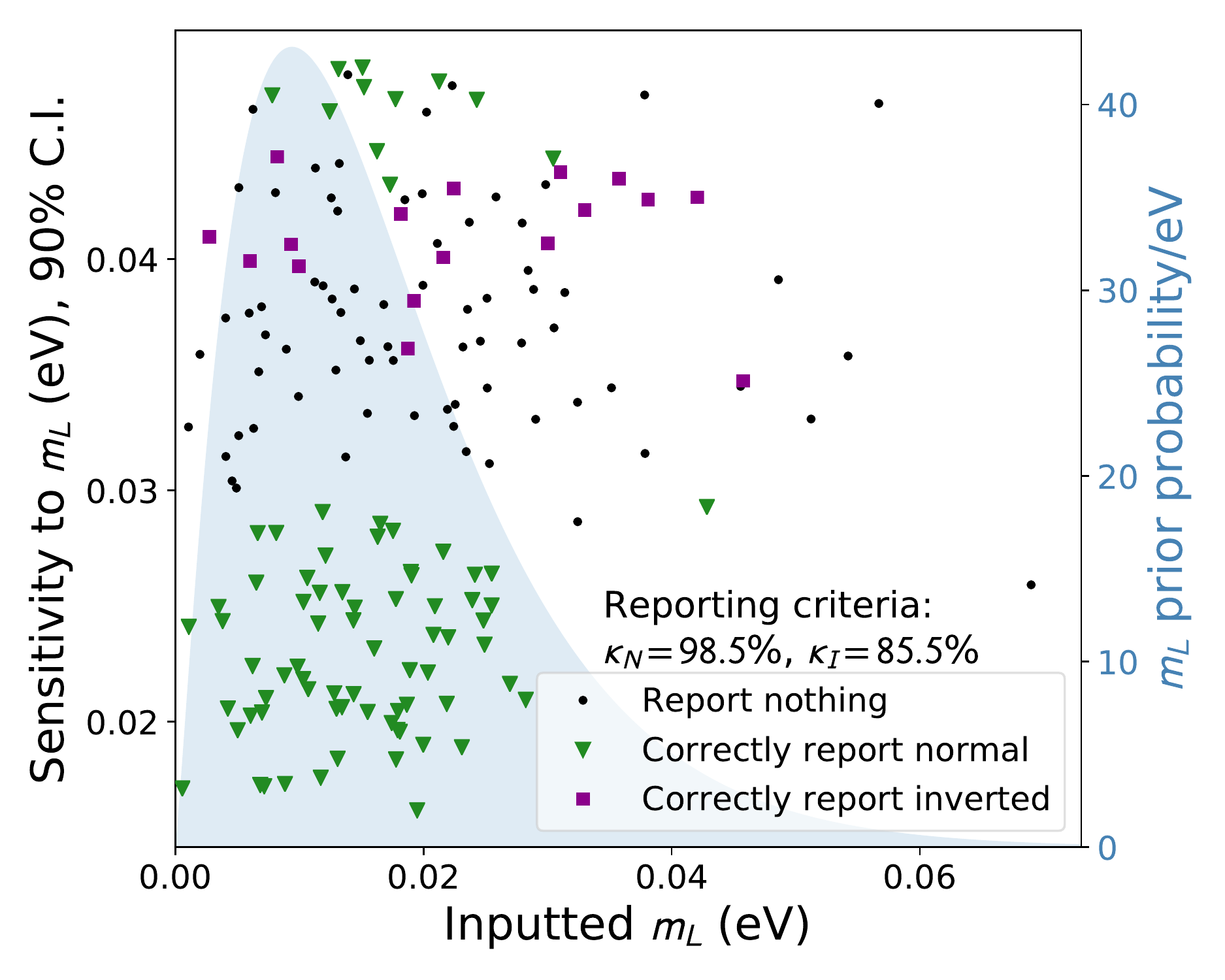} %3-17_MH-reporting_mL-sensitivity-vs-m_L_88_4yrs.pdf
\caption{Distribution of mass ordering results with respect to $m_L$ and mass sensitivity ($\Delta t=2$\,yrs). { The normal (inverted) ordering was reported when a 98.5\% (85.5\%) $\eta$ credible interval excluded one ordering and was consistent with the other.}}
\label{fig:MHreporting}
\end{figure}

For this study, we chose to employ a {\it two-neutrino} spectral model, as opposed to constraining the ordering based on a {\it single mass} measurement ---for which $m_\beta\lessapprox 48~\mathrm{meV}$ rules out an inverted ordering. While it would be impossible to resolve the inverted ordering using a one-neutrino model, a two-neutrino analysis can enable an  inverted ordering determination}. In other words, the process of inference is sensitive to fine structure near the endpoint of the spectrum produced by individual neutrino mass eigenstates. 
%This is true even for choices of reporting criteria that suppress false positive claims.

%By no means certain -- given some configurations of parameter values, according to our calibration scheme, there is not enough information to make an ordering claim.

%\vspace{0 pt}

\section{Conclusions}

In this paper, we presented a Bayesian approach to analyzing sensitivity to the neutrino mass scale and ordering. That approach included a calibration, {which quantified the performance of two processes: inferring information and reporting results.} Our sensitivity and calibration procedures are applicable to any experiment that produces information regarding the mass scale and ordering. {These procedures also serve as templates for sensitivity studies by other physics experiments---whether they measure continuous or discrete parameters. As design planning for Project 8's final phase advances,}   future work will include a detailed {analysis of }systematic features to inform more precise priors in {a Project 8-specific study.}

{Using the $\beta$ spectrum model developed here, and given the experimental expectations in Section~\ref{4a}, we find that a high-precision direct mass experiment could resolve the electron-weighted neutrino mass $m_\beta\approx m_1$ in a 90\% credible interval, with a ``true claim rate" or coverage of $(90.0\pm2.0)$\%. For very small $m_\beta$, the width of this interval approaches 40\,meV, and for $m_\beta>0.5\,$eV, the average width is only 5\,meV. A similar analysis may be employed to search for and measure the mass(es) of sterile neutrino states, each of which would produce one kink in the $\beta$ spectrum.}

{This study also investigates the tritium $\beta$-decay technique's sensitivity to the neutrino mass ordering. We emphasize that, by using a utility function to judge whether to report an ordering result, it is possible not only to predict the probability of a false ordering claim, but also to determine a reporting tolerance (here, the $\eta$ interval credibility) that minimizes the risk of false claims. For the experimental parameters assumed here and a two-year runtime, we would recommend reporting a normal ordering result when a 98.5\% posterior credible interval on the light-mass fraction $\eta$ contains $|U_{e1}|^2+|U_{e2}|^2$ but not $|U_{e3}|^2$. To report an inverted ordering determination, the opposite should hold for an 85.5\% interval around $\eta$. Those reporting criteria enable the normal (inverted) ordering to be resolved $\approx87$\% (22\%) of the time, with a $\approx0$\% false claim rate. It is also possible to infer posteriors on individual neutrino masses. When sensitivity to the lightest mass is better than 0.03\,eV, it is nearly always possible to resolve the mass ordering.}

These results demonstrate that we can access more information by modeling the full spectral shape {than would be possible using a one-neutrino model in terms of $m_\beta$.}
%simply identifying the energy where the spectrum ends. This is true in two key ways: First, a spectral shape analysis allows mass sensitivities to surpass the energy resolution at high statistics. 
As more events are detected, the spectral shape method becomes increasingly sensitive to count rate kinks that inform inferences about individual neutrino masses and their ordering. Direct mass experiments thus offer a unique potential probe of individual $|U_{ei}|$ matrix elements, complementary to oscillations-based probes of their products.
%Moreover, if one relaxes the condition that the fractions of the spectrum generated by each mass eigenstate sum to one, a $\beta$-decay measurement could constrain neutrino mixing unitarity. This would serve as an independent test of theories regarding the origin of flavor-mass state misalignment~\cite{Ellis2020}}.

%Given the experimental expectations detailed in Section~\ref{4a}, we find that a high-precision direct neutrino mass experiment could expect to resolve the mass ordering for various parameter  configurations. It is more often possible to report a ``normal ordering" result for lower neutrino masses and better mass sensitivity. Indeed, our simulated experiments consistently resolve the normal ordering for $m_L$ below $\approx\,0.04$\,eV, measured within credible intervals of $\approx0.02$\,eV. Furthermore, we find that we can access more information by modeling and analyzing the full spectral shape than would be possible by simply identifying the energy where the spectrum ends. This is true in two key ways: First, a spectral shape analysis allows mass sensitivities to surpass the energy resolution at high statistics. Second, as more events are detected, the spectral shape method becomes increasingly sensitive to a count rate ``kink'' that informs inferences about individual neutrino masses and their ordering.
%Our two-neutrino spectral shape analysis makes it possible  The non-zero inverted ordering reporting rate is made possible by our 

\section*{Acknowledgments}

The authors would like to thank Andr\'{e} de Gouv\^{e}a for insight and discussions.
This material is based upon work supported by the following sources: the U.S. Department of Energy Office of Science, Office of Nuclear Physics, under Award No.~DE-SC0020433 to Case Western Reserve University (CWRU), under Award No.~DE-SC0011091 to the Massachusetts Institute of Technology (MIT), under the Early Career Research Program to Pacific Northwest National Laboratory (PNNL), a multiprogram national laboratory operated by Battelle for the U.S. Department of Energy under Contract No.~DE-AC05-76RL01830, under Early Career Award No.~DE-SC0019088 to Pennsylvania State University, under Award No.~DE-FG02-97ER41020 to the University of Washington, and under Award No.~DE-SC0012654 to Yale University; the National Science Foundation under Award Nos.~PHY-1205100 to MIT; the Cluster of Excellence “Precision Physics, Fundamental Interactions, and Structure of Matter” (PRISMA+ EXC 2118/1) funded by the German Research Foundation (DFG) within the German Excellence Strategy (Project ID 39083149); the Laboratory Directed Research and Development (LDRD) 18-ERD-028 at Lawrence Livermore National Laboratory (LLNL), prepared by LLNL under Contract DE-AC52-07NA27344, LLNL-JRNL-817667; the LDRD program at PNNL; the University of Washington Royalty Research Foundation; Yale University; and the Karlsruhe Institute of Technology (KIT) Center Elementary Particle and Astroparticle Physics (KCETA).  A portion of the research was performed using the Engaging cluster at the MGHPCC facility.

\bibliographystyle{apsrev4-1}
\bibliography{BayesMassHierarchy_RevisedAnalysis}

\appendix
\section{Approximate spectral model}\label{ApA}
{ The approximate $\beta$ spectral model for Bayesian inference in Eq.~\ref{eq:FofK} has a corresponding cumulative distribution function. It is given by
\begin{equation}
\begin{split}
&\mathcal{G}_i^{\text{CDF}}(K) = \int_{K}^{\infty}\mathcal{F}_i(K')dK' = \\
&\Big[ G_\text{A}(K | m_i, Q_T, \sigma) - G_\text{B}(K | m_i, Q_T, \sigma, K_{\text{min}}) \Big ]/C
\end{split}
\nonumber
\end{equation}
\noindent where
\begin{equation}
\nonumber
\begin{split}
G_\text{A} &= \mathcal{N}(Q_T-K|m_i, \sigma)2\sigma^2\cdot\Big[4 \sigma^2  - m_i^2 + 2 m_i (Q_T - K) \\
&+ 2 (Q_T - K)^2\Big]
+ \text{Erfc}\Bigg(\frac{m_i - Q_T + K}{\sqrt{2}\sigma}\Bigg)\\
&\times\Big[m_i^3 + (Q_T - K) \cdot\Big(6 \sigma^2 - 3 m_i^2 + 2 (Q_T - K)^2\Big)\Big]
\end{split}
\end{equation}
\begin{equation}
\nonumber
\begin{split}
G_\text{B} &= \mathcal{N}(Q_T-K|Q_T - K_{\text{min}}, \sigma)2\sigma^2\cdot\Big[4 \sigma^2 - 3m_i^2\\
&+ 2\Big((Q_T - K_{\text{min}})^2
- (Q_T - K_{\text{min}})(Q_T - K) + (Q_T - K)^2\Big)\Big]\\
&+ \text{Erfc}\Bigg(\frac{K-K_{\text{min}}}{\sqrt{2}\sigma}\Bigg) \Big[(Q_T-K_{\text{min}})\Big(3m_i^2-2(Q_T-K_{\text{min}})^2\Big)\\ 
&+ (Q_T - K)(6 \sigma^2 - 3 m_i^2 + 2 (Q_T - K)^2)\Big] \\
C &= \Big [ G_\text{high}(K) - G_\text{low}(K) \Big ] \Big |_{0}^{\infty}.
\end{split}
\end{equation}
\\
\noindent We implemented this function in Stan and employed it to analyze fake spectra.}

\section{\label{ApB} Prior distribution definitions}
The prior distributions used in this paper are defined as follows. Each distribution is implemented via a Stan function that outputs the log of the probability density of a parameter $y$~\cite{Stanual2020}.

\begin{enumerate}
\item Normal distribution:
\begin{equation}
\quad \quad \mathcal{N}(\mu, \sigma) \equiv \mathcal{N}(y | \mu, \sigma) = \frac{1}{\sqrt{2\pi}\sigma}\text{exp}\Bigg(-\frac{1}{2}\Big(\frac{y-\mu}{\sigma} \Big)^2\Bigg)
\nonumber
\end{equation}

\item Gamma distribution:
\begin{equation}
\begin{split}
\ \, \gamma(\alpha, \beta) \equiv \ &\gamma(y | \alpha, \beta) = \frac{\beta^\alpha}{\Gamma(\alpha)} y^{\alpha-1}\text{exp}(-\beta y) \\
\text{where} \ &\Gamma(\alpha) = \int_0^{\infty} x^{\alpha-1} e^{-x} dx
\end{split}
\nonumber
\end{equation}

\item Log-normal distribution:
\begin{equation}
\begin{split}
\quad \ \, \text{lognorm}(\mu, \sigma) &\equiv \text{lognorm}(y | \mu, \sigma) \\ &= \frac{1}{\sqrt{2\pi}\sigma y}\text{exp}\Bigg(-\frac{1}{2}\Big(\frac{\text{log}y-\mu}{\sigma} \Big)^2\Bigg)
\end{split}
\nonumber
\end{equation}

\end{enumerate}

%{\section*{Appendix C}\label{ApC}
%The performance of a likelihood model given data can be evaluated on two axes: how informative the model is, and how accurate the results are. ... shrinkage and z-score metrics.}

\end{document}